\input epsf
\input harvmac
\input srctex.sty

\def\tilb{\beta^{^{\hskip -.15cm \sim}}}

\def\K3{{\bf K3}}
\def\journal#1&#2(#3){\unskip, \sl #1\ \bf #2 \rm(19#3) }
\def\andjournal#1&#2(#3){\sl #1~\bf #2 \rm (19#3) }

\def\tilde{\widetilde}

\def\frac#1#2{{#1\over#2}}

\def\inbar{\,\vrule height1.5ex width.4pt depth0pt}
\def\IC{\relax\hbox{$\inbar\kern-.3em{\rm C}$}}
\def\IR{\relax{\rm I\kern-.18em R}}
\def\IP{\relax{\rm I\kern-.18em P}}

\catcode`\@=11
\def\slash#1{\mathord{\mathpalette\c@ncel{#1}}}
\overfullrule=0pt

\def\BB{{\cal B}}
\def\CC{{\cal C}}

\def\OO{{\cal O}}

\def\ZZ{{\cal Z}}

\def\underrel#1\over#2{\mathrel{\mathop{\kern\z@#1}\limits_{#2}}}

\catcode`\@=12


%

\def \sinh{{\rm sinh}}

\def\exp{{\rm exp}}


\def\IL{\relax{\rm I\kern-.18em L}}
\def\IH{\relax{\rm I\kern-.18em H}}
\def\IR{\relax{\rm I\kern-.18em R}}
\def\IC{\relax\hbox{$\inbar\kern-.3em{\rm C}$}}
\def\IZ{{\bf Z}}





\def\makeblankbox#1#2{\hbox{\lower\dp0\vbox{\hidehrule{#1}{#2}%
   \kern -#1
   \hbox to \wd0{\hidevrule{#1}{#2}%
      \raise\ht0\vbox to #1{}
      \lower\dp0\vtop to #1{}
      \hfil\hidevrule{#2}{#1}}%
   \kern-#1\hidehrule{#2}{#1}}}%
}%
\def\hidehrule#1#2{\kern-#1\hrule height#1 depth#2 \kern-#2}%
\def\hidevrule#1#2{\kern-#1{\dimen0=#1\advance\dimen0 by #2\vrule
    width\dimen0}\kern-#2}%
\def\openbox{\ht0=1.2mm \dp0=1.2mm \wd0=2.4mm  \raise 2.75pt
\makeblankbox {.25pt} {.25pt}  }

\def\bun#1/#2{\leavevmode
   \kern.1em \raise .5ex \hbox{\the\scriptfont0 #1}%
   \kern-.1em $/$%
   \kern-.15em \lower .25ex \hbox{\the\scriptfont0 #2}%
}

\def\opensquare{\ht0=3.4mm \dp0=3.4mm \wd0=6.8mm  \raise 2.7pt
\makeblankbox {.25pt} {.25pt}  }


\def\sector#1#2{\ {\scriptstyle #1}\hskip 1mm
\mathop{\opensquare}\limits_{\lower
1mm\hbox{$\scriptstyle#2$}}\hskip 1mm}

\def\tsector#1#2{\ {\scriptstyle #1}\hskip 1mm
\mathop{\opensquare}\limits_{\lower
1mm\hbox{$\scriptstyle#2$}}^\sim\hskip 1mm}

\def \ov {\over}
\def \p {\partial}
\def \ha {{1 \ov 2}}
\def\le{\left}
\def\ri{\right}

\def\IZ{{\bf Z}}

\def \om {\omega}
\def \ep {\epsilon}

\def\De{\Delta}

\def\Om{\Omega}


\lref\HartleTP{
  J.~B.~Hartle and S.~W.~Hawking,
  Phys.\ Rev.\ D {\bf 13}, 2188 (1976).
}

\lref\siopsisN{
  S.~Musiri, S.~Ness and G.~Siopsis,
  ``Perturbative calculation of quasi-normal modes of AdS Schwarzschild  black
  holes,''
  Phys.\ Rev.\ D {\bf 73}, 064001 (2006)
  [arXiv:hep-th/0511113].
}

\lref\JimboSS{ M.~Jimbo and T.~Miwa, ``QKZ equation with $|$q$|$=1 and correlation functions of the
XXZ model in the gapless regime,'' J.\ Phys.\ A {\bf 29}, 2923 (1996) [arXiv:hep-th/9601135].
}

\lref\HerzogPC{ C.~P.~Herzog and D.~T.~Son, ``Schwinger-Keldysh propagators from AdS/CFT
correspondence,'' JHEP {\bf 0303}, 046 (2003) [arXiv:hep-th/0212072].
}

\lref\SonSD{ D.~T.~Son and A.~O.~Starinets, ``Minkowski-space correlators in AdS/CFT
correspondence: Recipe and applications,'' JHEP {\bf 0209}, 042 (2002) [arXiv:hep-th/0205051].
}

\lref\KlebanovTB{
  I.~R.~Klebanov and E.~Witten,
  ``AdS/CFT correspondence and symmetry breaking,''
  Nucl.\ Phys.\ B {\bf 556}, 89 (1999)
  [arXiv:hep-th/9905104].
}

\lref\MarolfFY{
  D.~Marolf,
  ``States and boundary terms: Subtleties of Lorentzian AdS/CFT,''
  arXiv:hep-th/0412032.
}

\lref\BanksDD{
  T.~Banks, M.~R.~Douglas, G.~T.~Horowitz and E.~J.~Martinec,
  ``AdS dynamics from conformal field theory,''
  arXiv:hep-th/9808016.
}

\lref\PolyaLatta{
  G.~Polya and G.~Latta,
  ``Complex Variables,''
  Wiley, New York 1974 ISBN: 0471693308.
}

\lref\BalasubramanianSN{
  V.~Balasubramanian, P.~Kraus and A.~E.~Lawrence,
  ``Bulk vs. boundary dynamics in anti-de Sitter spacetime,''
  Phys.\ Rev.\ D {\bf 59}, 046003 (1999)
  [arXiv:hep-th/9805171].
}

\lref\hawkingpage{
  S.~W.~Hawking and D.~N.~Page,
  ``Thermodynamics Of Black Holes In Anti-De Sitter Space,''
  Commun.\ Math.\ Phys.\  {\bf 87}, 577 (1983).
}



\lref\HorowitzPQ{
  G.~T.~Horowitz and S.~F.~Ross,
  ``Possible resolution of black hole singularities from large N gauge
  theory,''
  JHEP {\bf 9804}, 015 (1998)
  [arXiv:hep-th/9803085].
}


\lref\BalasubramanianZV{ V.~Balasubramanian and S.~F.~Ross, ``Holographic particle detection,''
Phys.\ Rev.\ D {\bf 61}, 044007 (2000) [arXiv:hep-th/9906226].
}

\lref\HemmingKD{
  S.~Hemming, E.~Keski-Vakkuri and P.~Kraus,
  ``Strings in the extended BTZ spacetime,''
  JHEP {\bf 0210}, 006 (2002)
  [arXiv:hep-th/0208003].
}

\lref\KaplanQE{
  J.~Kaplan,
  ``Extracting data from behind horizons with the AdS/CFT correspondence,''
  arXiv:hep-th/0402066.
}

\lref\WittenZW{
  E.~Witten,
  Adv.\ Theor.\ Math.\ Phys.\  {\bf 2}, 505 (1998)
  [arXiv:hep-th/9803131].
}

\lref\LoukoTP{ J.~Louko, D.~Marolf and S.~F.~Ross, ``On geodesic propagators and black hole
holography,'' Phys.\ Rev.\ D {\bf 62}, 044041 (2000) [arXiv:hep-th/0002111].
}

\lref\LeviCX{ T.~S.~Levi and S.~F.~Ross, ``Holography beyond the horizon and cosmic censorship,''
arXiv:hep-th/0304150.
}

\lref\BerryMount{
  M.~V.~Berry and K.~E.~Mount,
  Rept.\ Prog.\ Phys.\  {\bf 35}, 315 (1972).
}

\lref\KnollSchaeffer{
  J.~Knoll and R.~Schaeffer,
  Annals Phys.\  {\bf 97}, 307 (1976).
}

\lref\GregoryAN{ J.~P.~Gregory and S.~F.~Ross, ``Looking for event horizons using UV/IR
relations,'' Phys.\ Rev.\ D {\bf 63}, 104023 (2001) [arXiv:hep-th/0012135].
}

\lref\BalasubramanianZU{ V.~Balasubramanian and T.~S.~Levi, ``Beyond the veil: Inner horizon
instability and holography,'' arXiv:hep-th/0405048.
}

\lref\BrecherGN{
  D.~Brecher, J.~He and M.~Rozali,
  ``On charged black holes in anti-de Sitter space,''
  arXiv:hep-th/0410214.
}

\lref\starinets{ A.~Nunez and A.~O.~Starinets, ``AdS/CFT correspondence, quasinormal modes, and
thermal correlators in  N = 4 SYM,'' arXiv:hep-th/0302026.
}

\lref\kos{ P.~Kraus, H.~Ooguri and S.~Shenker, ``Inside the horizon with AdS/CFT,''
arXiv:hep-th/0212277.
}

\lref\GubserBC{ S.~S.~Gubser, I.~R.~Klebanov and A.~M.~Polyakov, ``Gauge theory correlators from
non-critical string theory,'' Phys.\ Lett.\ B {\bf 428}, 105 (1998) [arXiv:hep-th/9802109].
}

\lref\malda{ J.~M.~Maldacena, ``The large N limit of superconformal field theories and
supergravity,'' Adv.\ Theor.\ Math.\ Phys.\  {\bf 2}, 231 (1998) [Int.\ J.\ Theor.\ Phys.\  {\bf
38}, 1113 (1999)] [arXiv:hep-th/9711200].
}

\lref\witten{ E.~Witten, ``Anti-de Sitter space and holography,'' Adv.\ Theor.\ Math.\ Phys.\  {\bf
2}, 253 (1998) [arXiv:hep-th/9802150].
}

\lref\maldat{ J.~M.~Maldacena, ``Eternal black holes in Anti-de-Sitter,'' arXiv:hep-th/0106112.
}

\lref\witt{ E.~Witten, ``Anti-de Sitter space, thermal phase transition, and confinement in gauge
theories,'' Adv.\ Theor.\ Math.\ Phys.\  {\bf 2}, 505 (1998) [arXiv:hep-th/9803131].
}

\lref\sussWi{
  L.~Susskind and E.~Witten,
  ``The holographic bound in anti-de Sitter space,''
  arXiv:hep-th/9805114.
}

\lref\shenker{
  L.~Fidkowski, V.~Hubeny, M.~Kleban and S.~Shenker,
  ``The black hole singularity in AdS/CFT,''
  JHEP {\bf 0402}, 014 (2004)
  [arXiv:hep-th/0306170].
}

\lref\mot{
  L.~Motl and A.~Neitzke,
  ``Asymptotic black hole quasinormal frequencies,''
  Adv.\ Theor.\ Math.\ Phys.\  {\bf 7}, 307 (2003)
  [arXiv:hep-th/0301173].
}

\lref\ricar{
  V.~Cardoso, J.~Natario and R.~Schiappa,
  ``Asymptotic quasinormal frequencies for black holes in non-asymptotically
  flat spacetimes,''
  J.\ Math.\ Phys.\  {\bf 45}, 4698 (2004)
  [arXiv:hep-th/0403132].
}

\lref\FestucciaSA{
  G.~Festuccia and H.~Liu,
  arXiv:hep-th/0611098.
}

\lref\horomal{
  G.~T.~Horowitz and J.~Maldacena,
  ``The black hole final state,''
  JHEP {\bf 0402}, 008 (2004)
  [arXiv:hep-th/0310281].
}

\lref\bekenstein{
  J.~D.~Bekenstein and L.~Parker,
  ``Path Integral Evaluation Of Feynman Propagator In Curved Space-Time,''
  Phys.\ Rev.\ D {\bf 23}, 2850 (1981).
}

\lref\KonoplyaZU{
  R.~A.~Konoplya,
  Phys.\ Rev.\  D {\bf 66}, 044009 (2002)
  [arXiv:hep-th/0205142].
}

\lref\HertogRZ{
  T.~Hertog and G.~T.~Horowitz,
  ``Towards a big crunch dual,''
  JHEP {\bf 0407}, 073 (2004)
  [arXiv:hep-th/0406134].
}

\lref\HertogHU{
  T.~Hertog and G.~T.~Horowitz,
  ``Holographic description of AdS cosmologies,''
  arXiv:hep-th/0503071.
}

\lref\siopsis{
  G.~Siopsis,
  ``Large mass expansion of quasi-normal modes in AdS(5),''
  Phys.\ Lett.\ B {\bf 590}, 105 (2004)
  [arXiv:hep-th/0402083].
}

\lref\berryG{
  M.~V.~Berry,
  ``Infinitely many Stokes smoothings in the gamma
 function,'' Proc. \ Roy. \ Soc. \  Lond. \
A {\bf 434}, 465 (1991).}

\lref\HorowitzJD{
  G.~T.~Horowitz and V.~E.~Hubeny,
  Phys.\ Rev.\  D {\bf 62}, 024027 (2000)
  [arXiv:hep-th/9909056].
}

\lref\loOr{
  See e.g. G.~T.~Horowitz and A.~R.~Steif,
  ``Singular string solutions with nonsingular initial data,''
  Phys.\ Lett.\ B {\bf 258}, 91 (1991).
}

\lref\HamiltonJU{
  A.~Hamilton, D.~Kabat, G.~Lifschytz and D.~A.~Lowe,
  ``Local bulk operators in AdS/CFT: A boundary view of horizons and
  locality,''
  arXiv:hep-th/0506118.
}

\lref\newtonii{R.~G.~Newton, ``Scattering Theory of Waves and Particles,'' McGraw-Hill Education
(January 1967)}

\lref\newtoni{R.~G.~Newton, ``Scattering Theory of Waves and Particles,'' McGraw-Hill Education
(January 1967), Chapter 12 }

\lref\btz{
  M.~Banados, C.~Teitelboim and J.~Zanelli,
  ``The Black hole in three-dimensional space-time,''
  Phys.\ Rev.\ Lett.\  {\bf 69}, 1849 (1992)
  [arXiv:hep-th/9204099].
}

\lref\FidkowskiFC{
  L.~Fidkowski and S.~Shenker,
  ``D-brane instability as a large N phase transition,''
  arXiv:hep-th/0406086.
}

\lref\MotlHD{
  L.~Motl,
  Adv.\ Theor.\ Math.\ Phys.\  {\bf 6}, 1135 (2003)
  [arXiv:gr-qc/0212096].
}

\lref\MotlCD{
  L.~Motl and A.~Neitzke,
  Adv.\ Theor.\ Math.\ Phys.\  {\bf 7}, 307 (2003)
  [arXiv:hep-th/0301173].
}

\lref\birrel{ N.~D.~Birrel and P.~C.~W.~Davies, ``Quantum fields in curved space,'' Cambridge
Monographs on Mathematical Physics (1982)}

\lref\NatarioJD{
  J.~Natario and R.~Schiappa,
  ``On the classification of asymptotic quasinormal frequencies for
  d-dimensional black holes and quantum gravity,''
  Adv.\ Theor.\ Math.\ Phys.\  {\bf 8}, 1001 (2004)
  [arXiv:hep-th/0411267].
}

\lref\vijayetal{ V.~Balasubramanian, P.~Kraus, A.~E.~Lawrence and S.~P.~Trivedi,
  ``Holographic probes of anti-de Sitter space-times,''
  Phys.\ Rev.\ D {\bf 59}, 104021 (1999)
  [arXiv:hep-th/9808017].}

\lref\FestucciaPI{
  G.~Festuccia and H.~Liu,
  ``Excursions beyond the horizon: Black hole singularities in Yang-Mills
  theories. I,''
  JHEP {\bf 0604}, 044 (2006)
  [arXiv:hep-th/0506202].
}

\lref\suneeta{
  T.R.Govindarajan~ and V.~Suneeta,
  ``Quasi-normal modes of AdS black holes : A superpotential approach''
  [arXiv:hep-th/0007084].
}

\lref\KalyanaRamaZJ{
  S.~Kalyana Rama and B.~Sathiapalan,
  Mod.\ Phys.\ Lett.\  A {\bf 14}, 2635 (1999)
  [arXiv:hep-th/9905219].
}

\lref\SchutzKM{
  B.~F.~Schutz and C.~M.~Will,
  Astrophys.\ J.\ {\bf 291}, L33 (1985)
}

\lref\GuinnBN{
  J.~W.~Guinn, C.~M.~Will, Y.~Kojima and B.~F.~Schutz,
  Class.\ Quant.\ Grav.\  {\bf 7}, L47 (1990).
}

\lref\LawrenceZE{
  A.~Lawrence and A.~Sever,
  JHEP {\bf 0610}, 013 (2006)
  [arXiv:hep-th/0606022].
}

\lref\CardosoCJ{
  V.~Cardoso, R.~Konoplya and J.~P.~S.~Lemos,
  Phys.\ Rev.\  D {\bf 68}, 044024 (2003)
  [arXiv:gr-qc/0305037].
}

\lref\DanielssonFA{
  U.~H.~Danielsson, E.~Keski-Vakkuri and M.~Kruczenski,

  JHEP {\bf 0002}, 039 (2000)
  [arXiv:hep-th/9912209].
}

\lref\DanielssonZT{
  U.~H.~Danielsson, E.~Keski-Vakkuri and M.~Kruczenski,
  Nucl.\ Phys.\  B {\bf 563}, 279 (1999)
  [arXiv:hep-th/9905227].
}

\lref\GrainDG{
  J.~Grain and A.~Barrau,
  Nucl.\ Phys.\  B {\bf 742}, 253 (2006)
  [arXiv:hep-th/0603042].
}

\lref\BriganteGZ{
  M.~Brigante, H.~Liu, R.~C.~Myers, S.~Shenker and S.~Yaida,
  Phys.\ Rev.\ Lett.\  {\bf 100}, 191601 (2008)
  [arXiv:0802.3318 [hep-th]].
}

\lref\Leaver{
  E.~W.~Leaver,
  Class.\ Quantum\ Grav.\ {\bf 9}, 1643 (1992)}

\lref\BriganteNU{
  M.~Brigante, H.~Liu, R.~C.~Myers, S.~Shenker and S.~Yaida,
  Phys.\ Rev.\  D {\bf 77}, 126006 (2008)
  [arXiv:0712.0805 [hep-th]].
}

\lref\BirminghamPJ{
  D.~Birmingham, I.~Sachs and S.~N.~Solodukhin,
  Phys.\ Rev.\ Lett.\  {\bf 88}, 151301 (2002)
  [arXiv:hep-th/0112055].
}

\lref\KokkotasBD{
  K.~D.~Kokkotas and B.~G.~Schmidt,
  Living Rev.\ Rel.\  {\bf 2}, 2 (1999)
  [arXiv:gr-qc/9909058].
}

\lref\PadmanabhanFX{
  T.~Padmanabhan,
  Class.\ Quant.\ Grav.\  {\bf 21}, L1 (2004)
  [arXiv:gr-qc/0310027].
}

\lref\MirandaVB{
  A.~S.~Miranda, J.~Morgan and V.~T.~Zanchin,
  arXiv:0809.0297 [hep-th].
}

\lref\MirandaQX{
  A.~S.~Miranda and V.~T.~Zanchin,
  Phys.\ Rev.\  D {\bf 73}, 064034 (2006)
  [arXiv:gr-qc/0510066].
}

\lref\ChoudhuryWD{
  T.~R.~Choudhury and T.~Padmanabhan,
  Phys.\ Rev.\  D {\bf 69}, 064033 (2004)
  [arXiv:gr-qc/0311064].
}



\Title{\vbox{\baselineskip12pt \hbox{hep-th/yymmnnn}
\hbox{MIT-CTP-3995}
\hbox{SCIPP 08/11}
}}%
 {\vbox{\centerline{A Bohr-Sommerfeld quantization formula}
 \medskip
 \centerline{for quasinormal frequencies of AdS black holes}
 }}

\smallskip
\centerline{Guido Festuccia$^1$ and Hong Liu$^2$ }
\medskip

\centerline{\it  ${^1}$Santa Cruz Institute for Particle Physics} \centerline{\it
University of California Santa Cruz} \centerline{\it Santa Cruz,
California, 95064}

\medskip

\centerline{\it  ${^2}$Center for Theoretical Physics} \centerline{\it
Massachusetts Institute of Technology} \centerline{\it Cambridge,
Massachusetts, 02139}

\smallskip

\smallskip

\smallskip

\vglue .3cm

\medskip
\noindent

We derive a quantization formula of Bohr-Sommerfeld type for computing
quasinormal frequencies for scalar perturbations in
an AdS black hole in the limit of large scalar mass or spatial momentum.
We then apply the formula to find poles in retarded Green functions of boundary CFTs on  $\IR^{1,d-1}$ and $\IR \times S^{d-1}$.
We find that when the boundary theory is perturbed by an operator of dimension $\De \gg 1$, the relaxation time back to equilibrium is given at zero momentum by ${1 \ov \De \, \pi T } \ll {1 \ov \pi T}$. Turning on a large spatial momentum can significantly increase it.
For a generic scalar operator in a CFT on $\IR^{1,d-1}$, there exists a sequence of poles near the lightcone whose imaginary part scales with momentum as $p^{-{d-2 \ov d+2}}$ in the large momentum limit. For a CFT on a sphere $S^{d-1}$ we show that
the theory possesses a large number of long-lived quasiparticles whose imaginary part is exponentially small in momentum.

\medskip

\Date{November, 2008}


\bigskip



\newsec{Introduction}

Black hole quasinormal modes are damped oscillations which describe the evolution of small classical perturbations in a black hole background. These perturbations die off by falling into the horizon of the black hole or, in the case of an asymptotically flat spacetime, by escaping to infinity. The frequency of the oscillations therefore is not real but has an imaginary part describing their damping rate. Study of such small perturbations  are important for understanding the stability of a black hole and have  important applications
for gravitational wave astronomy (see e.g.~\refs{\KokkotasBD} for a review).

In the AdS/CFT correspondence~\refs{\malda,\witten,\GubserBC},
as a finite temperature boundary gauge theory is described
by a black hole in AdS~\refs{\WittenZW}, study of quasinormal modes for an AdS black hole  is important for understanding dynamics of strongly coupled gauge theories
at finite temperature~\refs{\KalyanaRamaZJ,\HorowitzJD}. 
More explicitly, black hole quasinormal modes correspond in the dual gauge theory to poles of thermal retarded Green functions in the complex frequency plane~\refs{\BirminghamPJ,\SonSD}. Thus the decay of the black hole quasinormal modes describes in the gauge theory how small perturbations of the thermal state relax back to equilibrium.
In particular, when the imaginary part of a quasi-normal frequency is smaller than the inverse temperature and the spacing between the neighboring modes, it describes a quasiparticle in the boundary gauge theory.

\ifig\poleS{Pole structure in the complex frequency ($\om$) plane of the thermal Wightman function of a typical operator in the boundary theory. The poles in the lower half plane are the quasi-normal poles for the retarded Green function. Those in the upper half plane are reflections of the ones in the lower half plane with respect to the real axis as follows from the standard relation between a retarded and Wightman function. Each pole line extends to infinity. The analytic behavior of the Wightman function in  different regions of the frequency plane reflects the bulk geometry in different regions of the black hole spacetime. For example the analytic behavior near $\om=0$ reflects the geometry near the black hole horizon.} {\epsfxsize=6cm \epsfbox{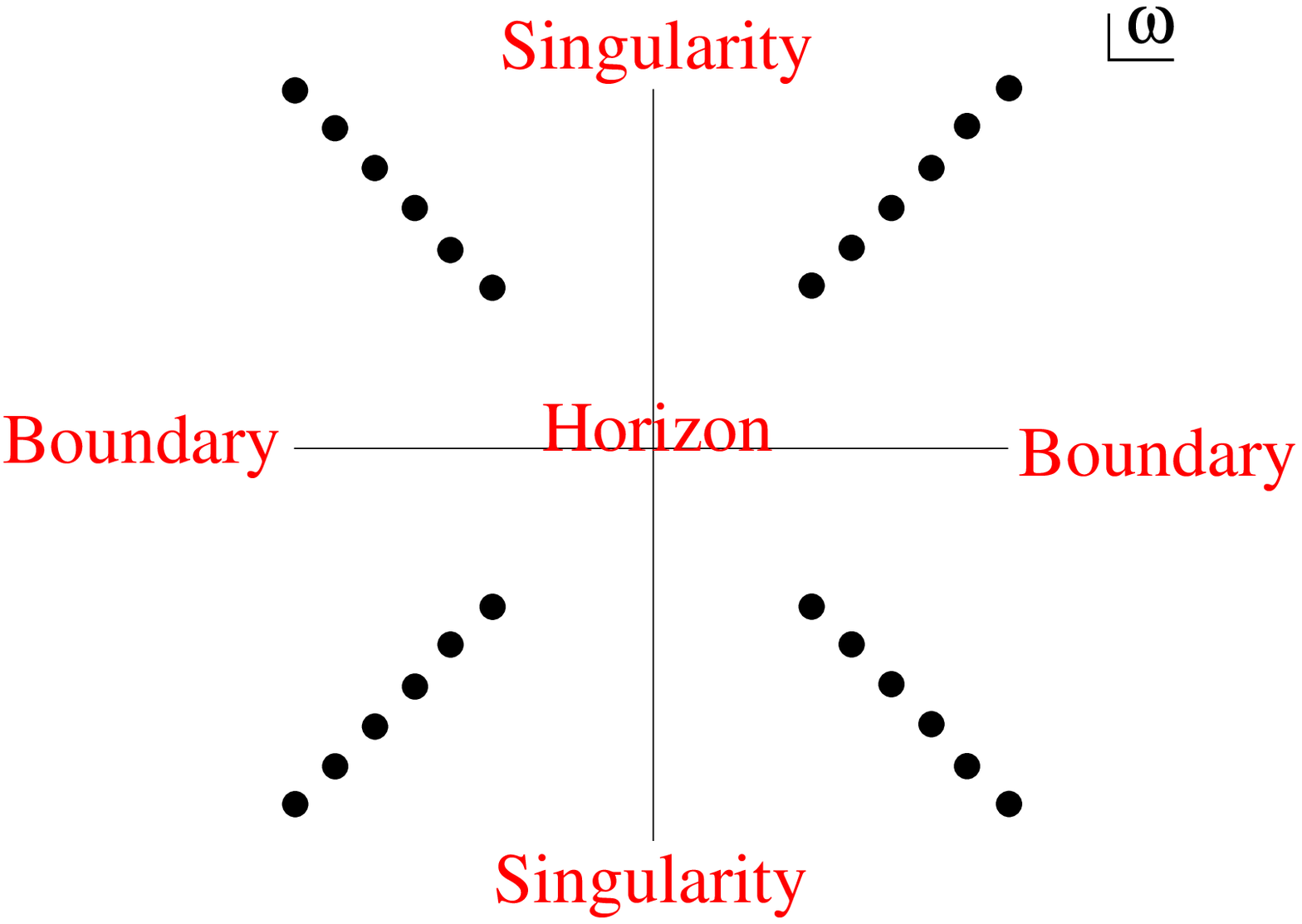}}

It has been argued in~\refs{\shenker,\FestucciaPI} that the analytic structure
of thermal correlation functions of the boundary theory encodes information about the region of spacetime behind the horizon and in particular the black hole singularities.
Understanding the structure of quasinormal poles\foot{Given the one-to-one correspondence between quasi-normal modes and poles of retarded correlators in the boundary, we will refer to poles in gauge theory correlators as quasi-normal poles to emphasize their
connection to the gravity side.} was an important element in the analysis.
For example, the thermal {\it Wightman} two-point function of a typical operator in the boundary theory has the pole structure shown in \poleS.
The four lines of poles divide the asymptotic frequency plane into four regions.
It was pointed out in~\refs{\FestucciaPI} that the analytic behavior of the Wightman function in  different regions of the frequency plane reflects the bulk geometry in different regions of the black hole spacetime, i.e. there exists a mapping between the
 complex-frequency plane of the boundary gauge theory  and the bulk geometry.
 Such a map can be considered as a generalization to regions inside the horizon of the standard IR/UV connection.
 The regions near the imaginary and real infinities in the frequency plane are mapped to the regions near the black hole singularities and the boundaries respectively, as labeled in~\poleS. In particular, the curvature divergence at the singularities is reflected in the divergences of derivatives of the YM response functions at large imaginary frequencies\foot{Such a map also indicates that the black hole singularities are resolved in quantum gravity, which corresponds to the boundary gauge theory at finite $N$ (the number color). See~\refs{\FestucciaPI} for a discussion.}.

Motivated from understanding the pole structure in \poleS\ at
 general spatial momenta, in this paper we derive\foot{The method was motivated from the approximation developed in~\refs{\FestucciaPI} to compute boundary correlation functions. } a general quantization formula of Bohr-Sommerfeld type for determining the quasinormal frequencies of scalar perturbations of an AdS black hole in the limit of large scalar mass or large spatial momentum. The basic idea for our approximation is as follows. We first rewrite the Laplace equation for the scalar field in terms of a  Schrodinger equation using the tortoise coordinate of the black hole geometry. We then
simplify the Schrodinger equation using a WKB approximation which can be considered as
a generalization (to finite or large masses) of the geometric optics approximation for massless modes. Quasinormal modes lie on anti-Stokes lines in the complex frequency plane where a certain subdominant correction becomes comparable to the leading WKB result and can be determined by requiring the two contributions to cancel each other exactly.

Compared to many other approximations already existing in the literature (see e.g.~\refs{\KokkotasBD} and~\refs{\GuinnBN\MotlHD\MotlCD\starinets\ricar\siopsis\suneeta\MirandaVB\MirandaQX\PadmanabhanFX\CardosoCJ\SchutzKM\KonoplyaZU\GrainDG\Leaver-\ChoudhuryWD} for an incomplete list), our method has the following features:

\item{$\bullet$} We introduce scaling limits which amount to a (generalized) geometric optics approximation. In the limits there is a natural connection between quasi-normal poles and complex geodesics of the black hole geometry as in~\refs{\FestucciaPI}.

\item{$\bullet$} The quasinormal frequencies are expressed elegantly in a quantization formula of Bohr-Sommerfeld type. We believe the formula should also apply to perturbations of a vector field or metric perturbations in a generic black hole geometry (e.g. charged) in AdS, once one has written the equation of motion of a given field in terms of a Schrodinger equation. Similar quantization formulas have been derived before in the case of asymptotically flat black holes starting with~\refs{\GuinnBN} and for AdS black holes in~\refs{\GrainDG}. Compared to these earlier work, our scaling limits
    introduce a small parameter and thus give the precise regimes to which the quantization formula can be reliably applied. This allows us to determine the location of quasinormal modes with very high frequency/overtone number which were not safely determined in previous work~\refs{\Leaver}. Our method can also be applied to determine the spectrum of quasinormal modes for any complex frequency value, including quasinormal pole lines starting far from the real axis.

\item{$\bullet$} The method applies well to quasinormal perturbations with large angular momentum (or momentum) which are hard to address using other methods.

\smallskip

 Applying our Bohr-Sommerfeld type formula to an AdS black hole geometry dual to a CFT$_d$ on $\IR^{1,d-1}$ or $\IR \times S^{d-1}$, we find the
 the following results among others: (we will quote the results for $d=4$, see main text for other dimensions)

\item{1.} For a scalar operator $\OO$ of dimension\foot{This corresponds to a bulk scalar field of mass $m$ which is related to $\De$ by $\De = {d \ov 2} + \sqrt{{d^2 \ov 4} + m^2 R^2}$, where $d$ is the spacetime dimension of the boundary theory and
     $R$ is the curvature radius of AdS.} $\Delta \gg 1$ in a CFT$_4$ on $\IR^{1,3}$, for momentum $p/T\ll \De$, the quasinormal poles of the corresponding bulk
    scalar field are given by
    \eqn\eokr{
    \om_n \approx (\pm 1 - i)\pi T (\De + 2 n -1 ) + O\le({1 \ov \De} \ri),  \qquad n=0,1,2,\cdots
    }
 where $T$ is the temperature. Equation \eokr\ implies when the thermal system is perturbed by the operator $\OO$, the relaxation back to equilibrium happens very fast, given by a time scale
    \eqn\rela{
    \tau_r \approx {1 \ov \De \pi T} \ll {1 \ov \pi T} \ .
    }
Also note that the spacing of the pole is independent of $\Delta$.
Similar results apply to a CFT on a sphere for  angular momentum not too large. For $ k \equiv {p \ov (\Delta-2) \pi T} \gg 1$  we instead find that
\eqn\rmmbj{
 \om \approx p \pm   e^{\mp {i \pi \ov 3}} (\pi T) \le(4 k \ri)^{-{1 \ov 3}}
  \le(\Delta -2+\sqrt{6} \le(2n +1 \ri) \ri), \quad n=0,1,\cdots
 }
Compared to \rela\ this implies that at large $k$, the relaxation time is enhanced  by a factor $k^{1\over 3}$.  Here the order of limit is important in deriving \rmmbj\ we take $\Delta \to \infty$ limit first and then $k$ large.

\item{2.} For any $\Delta$ (not necessarily large), in the limit $p \gg  \pi T$, the quasinormal poles can be written in $1/p$ expansion as
 \eqn\rnfNejh{
\om= p \pm  0.344 \, e^{\mp  {i\pi \over 3}} (\pi T)^{4\over 3}\, p^{-{1 \ov 3}} \, (\Delta-1 + 2n)^{4 \over 3},\;\;\;\; n=0,1, \cdots \ .
 }
Equations~\rmmbj\ and \rnfNejh\ have qualitatively similar behavior in their overlapping  region of applicability, but not exactly the same, implying the two limits of taking $\Delta$ and $p$ large do not commute. Note that for  \rnfNejh\ the spacings of the poles are not homogeneous. For $p \gg (\De-1)^4 \pi T$, the poles in~\rnfNejh\
 have an imaginary part which is much smaller than the temperature scale, but the imaginary parts are always of the same order as the spacings between the poles. Thus in the spectral function we expect not to see individual peaks for each pole in \rnfNejh, rather a broad
 peak close to the light cone $\om = p$ reflecting the effects of a large number of poles.
 In this sense \rnfNejh\ are more like a branch cut than quasi-particle poles.  Similar remarks also apply to~\rmmbj.

\item{3.} New interesting behavior arises for a theory on $S^{d-1}$ in the large angular momentum regime. Here one finds (as already noticed by~\refs{\GrainDG}) that for an operator of any dimension, when the angular momentum $l$ of a quasinormal mode is big enough, there exists a large number of quasinormal poles with a small imaginary part, proportional to $e^{-l (\cdots)}$.
    The spacing of the poles remains finite in the large $l$ limit. Thus we find
    a large number of long-lived quasi-particles with a large angular momentum\foot{Long-lived quasi-particles with a large momentum were also found in a CFT on $\IR^{1,3}$ dual to a Gauss-Bonnet gravity in~\refs{\BriganteGZ,\BriganteNU}.}.
    The existence of a large number of quasinormal modes with a very small imaginary part  follows from the following simple fact of the bulk geometry. For a black hole of a spherical horizon, there exists a stable circular orbit for a particle moving in its geometry when the angular momentum of the particle is big enough, since the centrifugal force prevents the particle from falling into the black hole. While classically the orbit is stable, in the quantum theory the particle can nevertheless tunnel into the black hole, giving rise to a small imaginary part.
    From the boundary theory point of view, however, the existence of such long-lived quasi-particles for a theory on $S^{d-1}$ is curious. We discuss their possible physical origins at end of section 6.

The organization of the paper is as follows. In section 2 we give a brief review of the geometry of a Schwarzschild black hole in AdS. In section 3 we rewrite the equation of motion in terms of a Schrodinger equation and discuss the condition for determining the quasinormal modes. In Section 4 we introduce a WKB approximation which can be considered as a generalized geometric optics approximation and present a quantization formula of Bohr-Sommerfeld type for finding the quasinormal modes.
We apply the formula in section 5 to a theory on $\IR^{d-1}$ and in section 6 to theories on $S^{d-1}$.  The detailed derivation of the Bohr-Sommerfeld type formula is given in Appendix A.

\newsec{Black hole geometry}

In this section we briefly review the classical geometry of a
Schwarzschild black hole in an AdS$_{d+1}$ ($ d \geq 3 $)
spacetime. The metric can be written as:
 \eqn\adsbh{
ds^2=-f(r)dt^2+f(r)^{-1}dr^2+r^2 d\Omega_{d-1}^2
 }
with
 \eqn\defF{
f(r)=r^2+1-{\mu \ov r^{d-2}}, 
 }
where $\mu$ is proportional to the mass of the black hole, and $d \Om_{d-1}^2$ denotes the metric on
a unit $(d-1)$-sphere. We have set the curvature radius of AdS to be unity, as we will do
for the rest of the paper. As $r \to \infty$, the metric goes over to that of global AdS. The fully extended black hole spacetime has two
disconnected time-like boundaries each of topology $S^{d-1} \times \IR$~(see Fig.~1 for the
Penrose diagram).
The event horizon radius $r_0$ is given by the unique positive
root of the equation
$$
 r^2+1-{\mu \ov r^{d-2}} = 0 \ ,
 $$
and the inverse Hawking temperature is given by
 \eqn\temper{
 \beta = {4 \pi \ov f'(r_0)} = {4 \pi r_0 \ov d r_0^2 + (d-2)} \ .
 }
When $\mu > 2$, \adsbh\ dominates the thermal ensemble and describes a boundary CFT$_d$ on $S^{d-1} \times \IR$ at a temperature given by $1/\beta$~\refs{\WittenZW}.

\ifig\penrose{Penrose diagram for the AdS black hole. A null
geodesic going from the boundary to the singularity is indicated
in the figure.} {\epsfxsize=4cm \epsfbox{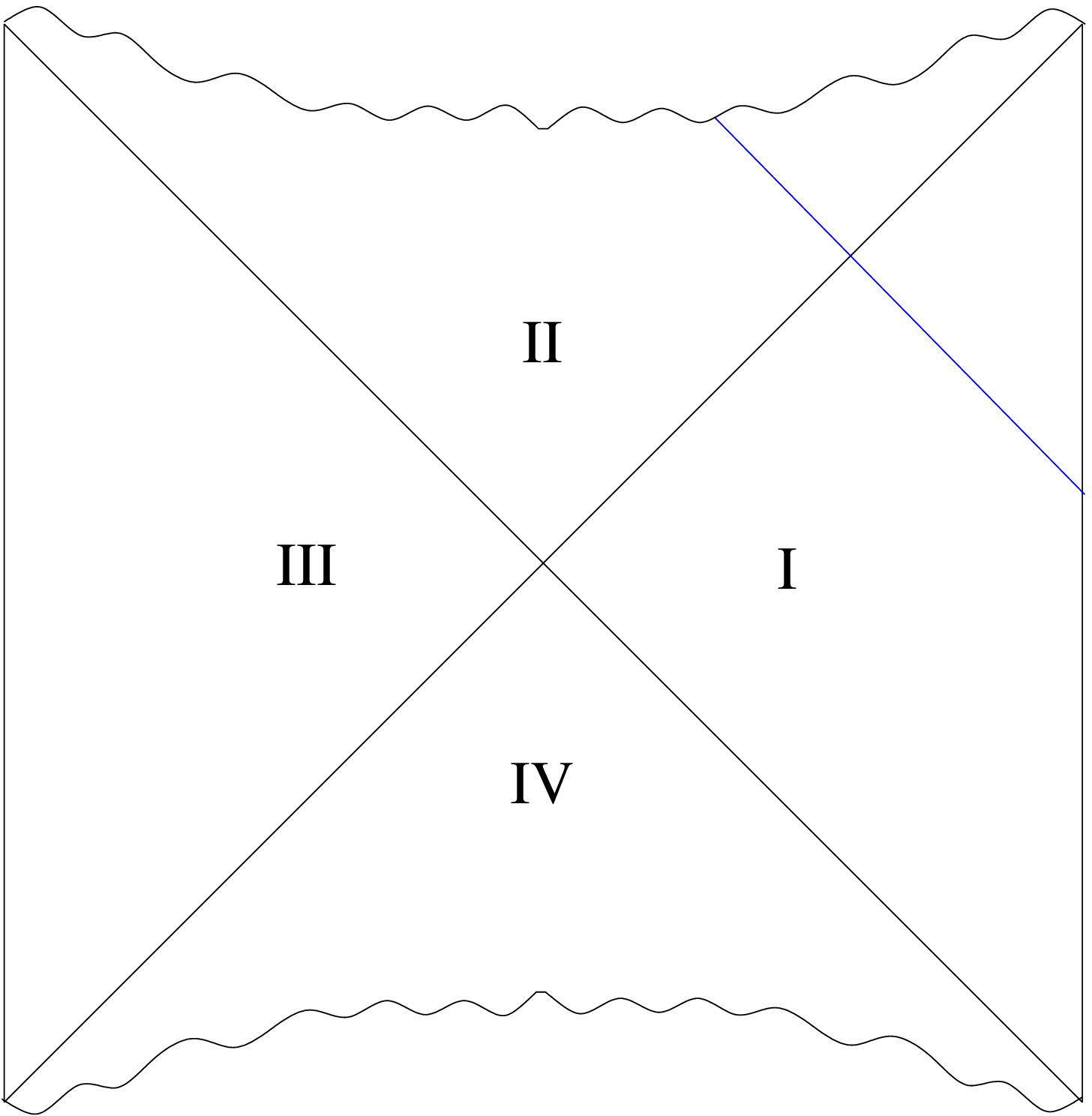}}

There also exists a black brane metric
 \eqn\bhmetric{
ds^2 = - f dt^2 + {1 \ov  f} dr^2 +  r^2 d x_i^2
 }
where $dx_i^2$ denotes the metric for a flat $(d-1)$-dimensional Euclidean space and
 \eqn\newf{
  f = r^2 - {1 \ov r^{d-2}} \ .
 }
The inverse Hawking temperature is
 \eqn\newtem{
\beta \equiv {1 \ov T} = {4 \pi \ov d} \ .
 }
The boundary manifold now consists of two copies of $\IR^{d,1}$. The metric \bhmetric\ describes a boundary CFT$_d$ on $\IR^{1,d-1}$ at a finite temperature given by \newtem. Note that for a CFT$_d$ on $\IR^{1,d-1}$, since the temperature is the only scale, it can be normalized at any given value. We choose \newtem\ for convenience. The boundary coordinates $(t, x_i)$ are dimensionless with basic unit give by ${4 \pi T \ov d}$.
Note that \bhmetric\ can be obtained from \adsbh\ by taking the following scaling limit
 \eqn\scalU{
r_0 \to \infty , \qquad  r \to r_0 r, \qquad t \to t/r_0, \qquad r_0^2 d \Om_{d-1}^2 =
 dx_i^2 \ .
 }
This is the high temperature limit of the boundary CFT on $S^{d-1}$. Equivalently we can fix the temperature scale as in \newtem, then effectively the size of the sphere goes to infinity and we get a theory on $\IR^{d-1}$.

To describe the black hole geometry it is often convenient to use
the tortoise coordinate:
 \eqn\tortoise{
 z (r) =  \int_{r}^\infty {dr \ov f(r)} =
 - \sum_{i=0}^{d-1} {1 \ov f'(r_i)} \log (r-r_i)
 \ ,
 }
where $r_i$ are zeros of $f$ with $r_0$ being the horizon.
The region outside the horizon (region I) corresponds to $z \in (0, +\infty)$. At the boundary $r
\to\infty$ we have $z \approx {1 \ov r} \to 0$. At the horizon $r \to r_0$ we have $z \approx -
{\beta \ov 4 \pi} \log (r-r_0) \to
 +\infty$.

\newsec{Propagation of a scalar field in an $AdS_{d+1}$Schwarzschild black hole}

Consider a scalar field\foot{Since the background Ricci scalar is a constant,  the $m^2$ term
below should be considered as the sum of the standard mass term and the coupling to the background
curvature.} with quadratic action
 \eqn\scalag{
S = - \ha \int dr d^{d} x \, \sqrt{- g} \, \left[ (\p \phi)^2 +
m^2 \phi^2 \right]
 }
in the background of \adsbh\ or \bhmetric. According to the standard AdS/CFT dictionary, the dimension $\De$ of the boundary theory operator $\OO$  dual to $\phi$ is  given by
\eqn\dimD{
 \De = {d \ov 2} + \nu, \qquad \nu = \sqrt{m^2 + {d^2 \ov 4}} \ .
 }

For \bhmetric\ let
 \eqn\eujs{
\phi = e^{- i \om t} e^{i \vec p \cdot \vec x} r^{-{d-1 \ov 2}}
\psi (\om, p;r),
 }
the Laplace equation for $\phi$ can then be written in terms of
the tortoise coordinate $z$ \tortoise\ as
 \eqn\TeomD{ \eqalign{
  \left(-  \p_z^2   + V_p (z) - \om^2 \right) \psi =0,
  }}
where $V_p$ is an implicit function of $z$ (below $p^2 = \vec p
\cdot \vec p$)
 \eqn\poeV{
 V_p(z) = f(r) \left[ { p^2 \ov r^2} +\nu^2 -{1 \ov 4}
 + {(d-1)^2 \ov 4 r^d} \right] \ .
 }
For \adsbh\ one replaces the plane wave in the $\vec x$ directions in \eujs\ by spherical harmonics
on $S^{d-1}$ and get \TeomD\ with now the potential given by
 \eqn\potenP{
V_l (z)=f(r)\le({(2 l + {d-2})^2 - {1} \ov 4 r^2}+\nu^2-{1 \ov 4 }+{\mu (d-1)^2 \ov 4 r^d } \ri)
 }
where $l$ is the angular momentum on $S^{d-1}$.

As discussed below \tortoise, the region outside the horizon
corresponds to $z \in (0, +\infty)$ with $z=0$ at the boundary and
$z \to + \infty$ at the horizon. Both \poeV\ and \potenP\ behave
near the boundary as
 \eqn\bougd{
V_p \approx {\nu^2 - {1 \ov 4} \ov z^2}, \qquad z \to 0 \ ,
 }
and near the horizon
 \eqn\fallho{
V_p  \propto e^{-{4 \pi \ov \beta} z } \to 0 , \qquad z \to
+\infty \ .
 }
The fact that for ${\rm Re} z \gg 1$, $r$ is a one-to one periodic
function of $z$ with a period $i {\beta \ov 2}$ implies that $V_p
(z)$ can be expanded for large real $z$ as
 \eqn\djsnV{
 V_ p (z) = \sum_{n=1}^\infty a_n e^{-{ 4 \pi n \ov \beta} z} \ .
 }
This property will be important in our discussion below.

\ifig\wkbj{Plots of the potential \poeV\ for $k=6,\;\nu=2$  (left) and \potenP\ with $l=6,\;\mu={5\ov 16},\;\nu=2$ (right).} {\epsfxsize=6cm
\epsfbox{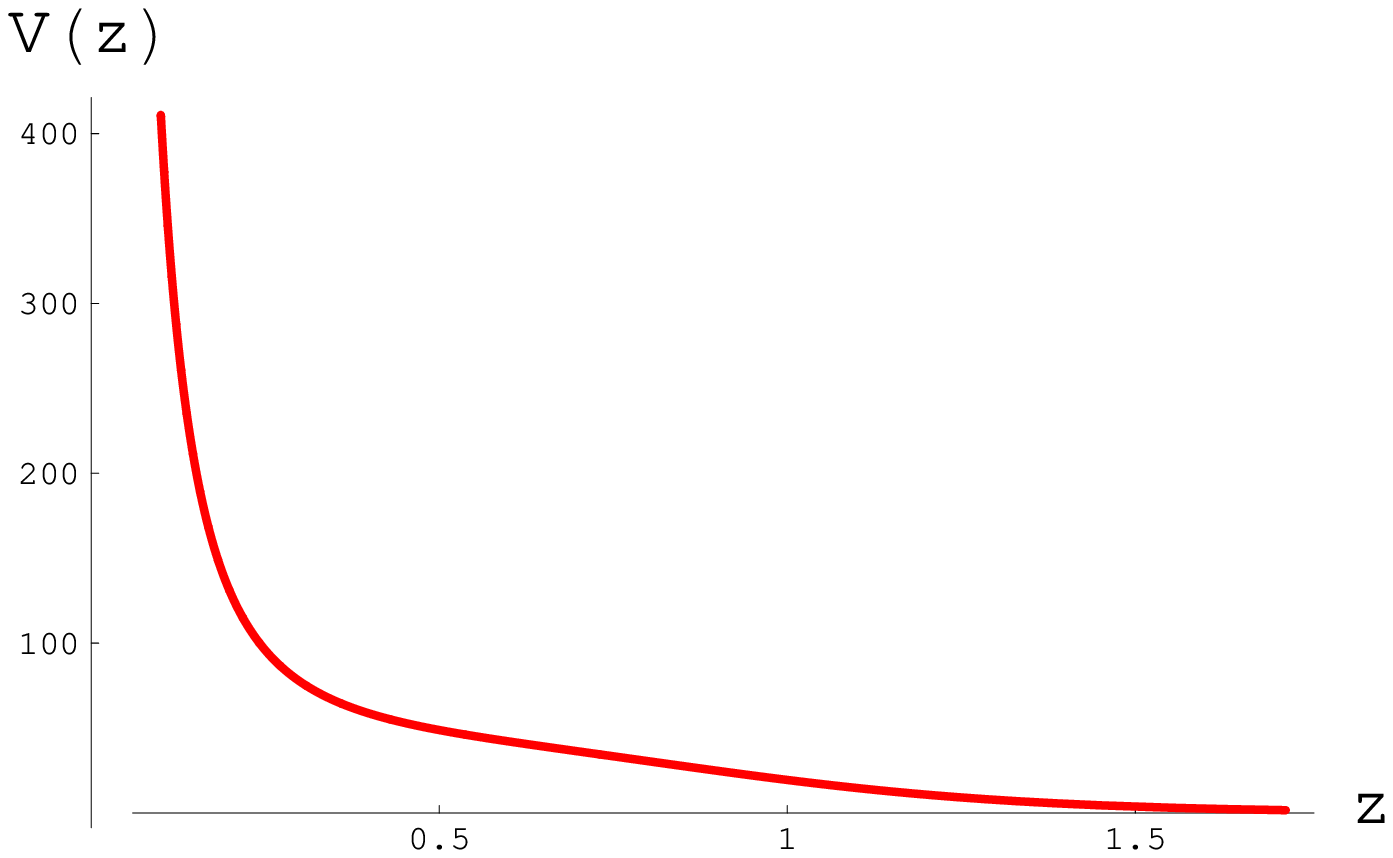} \epsfxsize=6cm \epsfbox{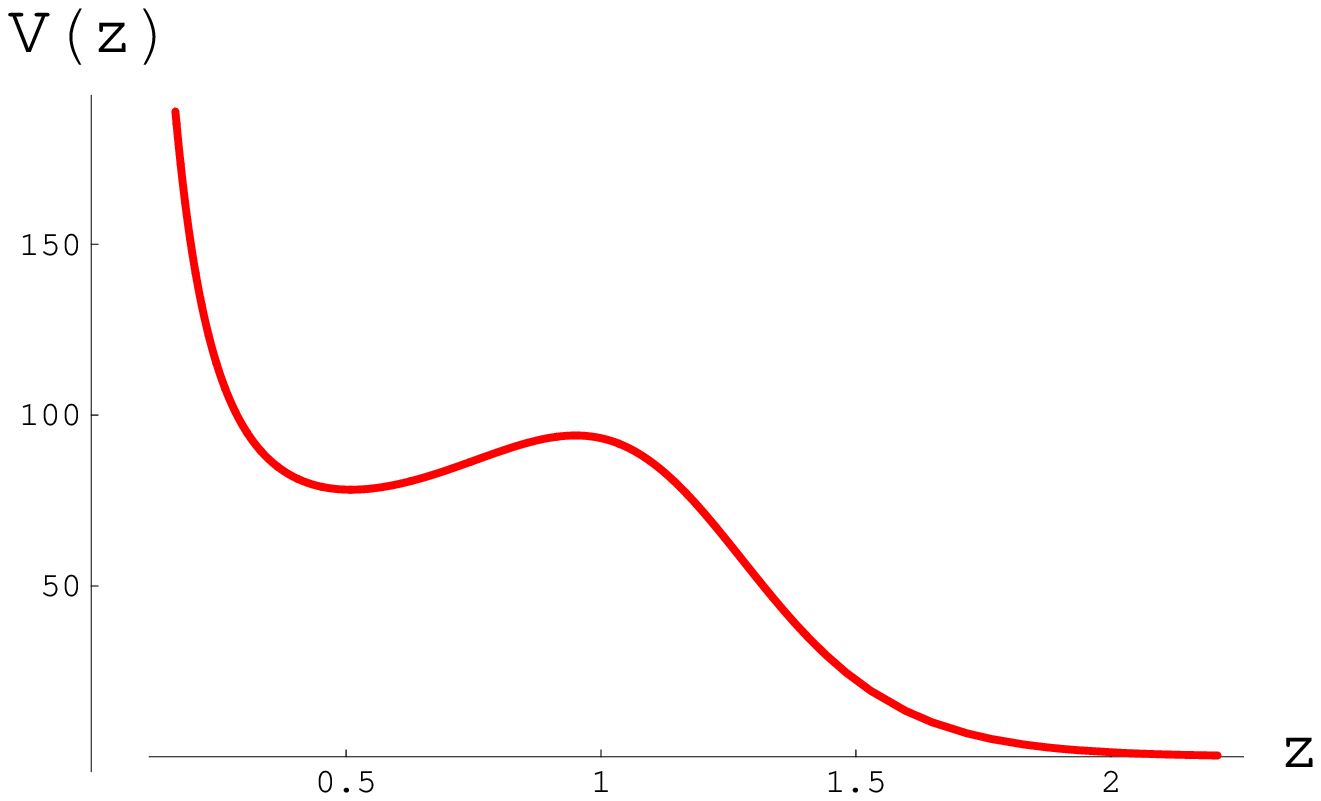}}

 Note that the potential \poeV\ is positive definite and
monotonic in $z \in (0, +\infty)$ for $\nu^2 > {1 \ov 4}$ and any $p^2 > 0$ (see \wkbj). The
potential \potenP\ is monotonic as in \wkbj\ for $l$ smaller than a critical value $l_c$, while for
$l > l_c$ the potential develops a well as in \wkbj . $l_c$ can be found by solving $V'(r) = V'' (r) =
0$. Its explicit value is rather complicated and we do not give it here. An implicit expression in
the large $\nu$ limit will be given in sec 7.2. The potential well reflects the fact that when the angular
momentum is sufficiently large there exist stable orbits for a particle moving outside the horizon.

In our discussions below, we will use the notation appropriate for
\bhmetric. The discussions apply to both \bhmetric\ and \adsbh\
unless mentioned explicitly.

For any given real $\om $, the Schrodinger equation \TeomD\ has a
unique normalizable mode $g_{\om, p}$, which we will normalize near the boundary as
 \eqn\Bonh{
 g_{\om p}(z)\approx z^{\ha +\nu}, \qquad z\rightarrow 0 .}
The asymptotic behavior of $g_{\om p}(z)$ near the horizon ($z\rightarrow \infty$) can be written as
\eqn\hort{g_{\om p}(z)\approx {1\ov 2 i \om}\le(f(-\om,p)e^{i \om z}-f(\om,p)e^{-i \om z}\ri),\qquad z\rightarrow \infty }
which defines the function $f (\om, p)$. Since the boundary condition \Bonh\ is real and independent of $\om, p$, for general complex values of $\om, p$, we have
 \eqn\ekkr{
f^* (\om,p) = f (-\om^*, p^*) \ .
}
When plugged back into \eujs\ the first term
in \hort\ describes a plane wave falling into  the horizon at $t \to + \infty$, while the second term describes a wave coming out of  the horizon at $t \to -\infty$.

At a given $p$, quasinormal frequencies for $\phi$ are those (complex) values of $\omega$ for which the {\it normalizable} solution is in-falling at the horizon, i.e. the solution behaves as
$e^{i \om z}$ for $z\rightarrow \infty$. From \hort, the quasinormal frequencies $\om_q (p)$ thus satisfy
 \eqn\quen{
 f(\om_q ,p)=0 \ .
 }
One can show that if the potential \poeV\ does not admit bound states (which is true for real momenta) the only zeroes of $f(\om,p)$ are in the lower half $\om$-plane. Also for real $p$ it follows from \ekkr\ that if $\om_q$ is a quasi-normal frequency, so is $-\om_q^*$, i.e. they are symmetric with respect to the lower imaginary axis in the complex $\om$-plane.

Using the prescription for computing the retarded Green function in AdS/CFT~\refs{\SonSD}
the quasi-normal frequencies are precisely the poles of the momentum-space retarded Green function $G_R (\om, p)$ for the operator dual to $\phi$ in the boundary gauge theory.
 In particular, those whose imaginary part is much smaller then the real part correspond to quasi-particles in the boundary theory.

\newsec{Quasi-normal modes from WKB approximation}

We now develop an approximate method to determine the function $g_{\om p}(z)$, from which
we obtain the quasi-normal modes which are zeros of $f(\om,p)$ defined in~\hort.
This method is applicable to values of the momentum $p$ of the same order as $\om$ in contrast to other approximations considered in the literature\refs{\ricar,\siopsis} which require $\om\gg p$. We first introduce the approximation and then summarize our results for the quasinormal frequencies, leaving the details of their derivation to Appendix~A.
For simplicity of notations, below we will use the notation appropriate for the flat case \poeV. The discussion applies without change to the sphere case \potenP.

\subsec{Large mass limit}

Consider the following large $\nu$ limit with $u$
and $\vec k$ fixed
 \eqn\largnu{
 \om = \nu u, \qquad \vec p = \nu  \vec k, \qquad \nu \gg  1 \ ,
 }
i.e. we take the mass $m$ of $\phi$ to be large and ``measure'' the frequency $\om$ and momentum $\vec p$ in units of $m$. This is analogous to a geometric optics limit in which we take $\om$ and $k$ large but keep the velocities fixed. In this limit, one can solve \TeomD\ approximately using the standard WKB
method with $1/\nu$ playing the role of $\hbar$. More explicitly, writing $\psi = e^{\nu S}$
equation \TeomD\ becomes
 \eqn\liuoG{
 - (\p_z S)^2 - {1 \ov \nu} \p_z^2 S + V(z)  +
 {1 \ov \nu^2} Q(z) = u^2
 }
with
 \eqn\degFV{
 V(z) = f \le(1 + {k^2 \ov r^2} \ri) \
 }
and
 \eqn\ghdr{
Q(z) =  f \le({(d-1)^2 \ov 4 r^d} -{1 \ov 4}\ri) \ .
 }

\ifig\potd{Schematic plot of the potential \degFV\ for $\vec k\in R$. The boundary is at $z=0$ the
horizon at $z=\infty$.} {\epsfxsize=6cm \epsfbox{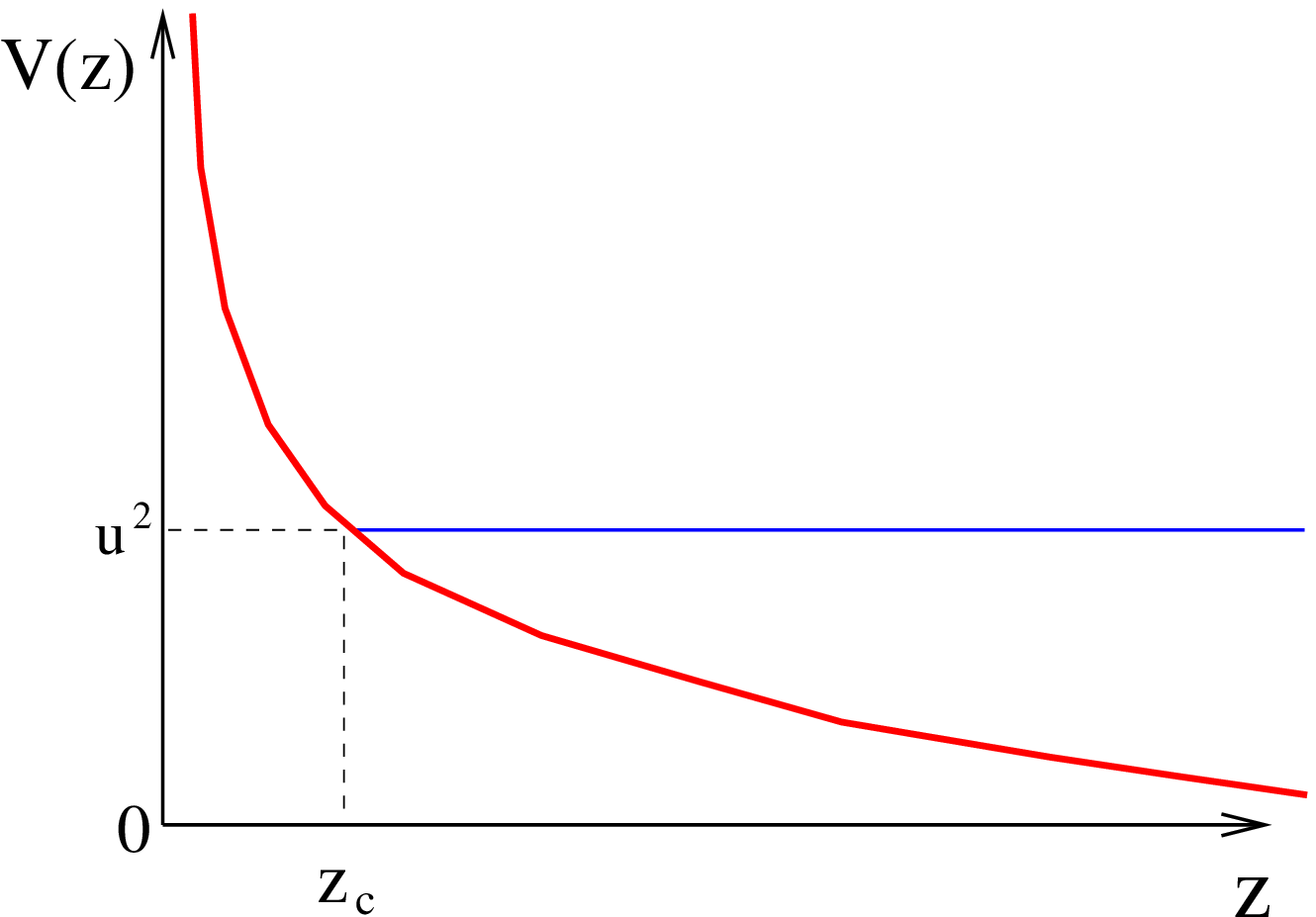}}

We first restrict to real $\vec k $, for which case the potential \degFV\ (see \potd ) is a
monotonically decreasing function for $z \in (0, +\infty)$, and consider positive energy scattering
sates with $u
>0$. Solving \liuoG\ order by order in $1/\nu$, we find that in the classically
forbidden region (i.e. for $z<z_c$ in \potd) the exponentially
decreasing solution to \TeomD\ can be written as
 \eqn\splg{
 g_{\om p}^{(wkb)} (z)
 ={A \ov \sqrt{\kappa (z)}}e^{ \nu \ZZ
 } \le( 1 + O\le(\nu^{-1} \ri) \ri)
 }
with\foot{The branch cuts for $\kappa (z)$ on the complex
$z$-plane are chosen so that they do not intersect the integration
contour in $\ZZ$.}
 \eqn\zzdef{\eqalign{
 \ZZ (z) & =  \int_{z_c}^z dz' \, \kappa (z') , \cr
  \kappa (z) & =  \sqrt{V(r) - u^2},\cr
  A&= \lim_{z\rightarrow 0}e^{ \nu(\log(z)- \ZZ(z))
 } \ .
 }}
$z_c = z(r_c)$ in the lower integration limit of \zzdef\ is the
turning point with $r_c$ given by the real positive root of the
equation
 \eqn\turnPw{
 V(r) = f (r) \le(1 + {k^2 \ov r^2} \ri) = u^2  \ .
 }
For $k^2,u^2>0$, equation \turnPw\ has a unique positive root $
r_c >1$. $\ZZ (z)$ satisfies the equation
 \eqn\wkbeR{\eqalign{
 &  {1 \ov f} \ZZ'^2 - {k^2 \ov r^2}  + {1 \ov f} u^2 = 1   \  \cr
 }}
with $\ZZ'(z_c) =0$. Note that to the order indicated in \splg,
the term in \liuoG\ proportional to $Q(z)$ does not contribute.
As discussed in detail in~\refs{\FestucciaPI} in a classically forbidden (allowed) region equation \wkbeR\ is simply the geodesic
equation for a space-like (time-like) geodesic.

The expression for $g^{(wkb)}_{\om p}$ in the classically
allowed region of the potential \degFV\ (i.e. for $z_c < z <
\infty$) follows from the standard connection formula
 \eqn\shde{
g_{\om p}^{(wkb)} ={2 A \ov  \sqrt{p (z)}} \cos \le(  \nu W
 - {\pi \ov 4} \ri) \le( 1 +
O\le(\nu^{-1} \ri) \ri)
 }
with
 \eqn\wdefB{
W= \int_{z_c}^{z} dz' \, p (z')  , \qquad p(z) = \sqrt{u^2 - f
\le(1 + {k^2 \ov r^2} \ri) }
 }
Near the event horizon \shde\ has the form
 \eqn\wekN{
g_{\om p}^{(wkb)} ={A \ov  \sqrt{u}} e^{i \om z + i
\delta_{\om} }  \le( 1 + O\le(\nu^{-1} \ri) \ri)
 + c.c.
 }

Higher order $1/\nu$ corrections in \splg\ and \shde\ can also be obtained
from \liuoG\ using the standard WKB procedure. In particular, the
term proportional to $Q(z)$ will be important at order $\nu^{-1}$.

\subsec{Large momentum limit}

For a scalar field with a mass $\nu \sim O(1)$, it is possible to use WKB methods to study \TeomD\ (or \potenP) in the large momentum (or angular momentum) limit $p\rightarrow \infty$ with $\nu$ fixed. Define
\eqn\defchi{
\om = p  u
}
then for $p\rightarrow \infty$ equation \TeomD\ becomes for $\psi=e^{p S}$:
\eqn\liuoGl{
- (\p_z S)^2 - {1 \ov p} \p_z^2 S + V(z)  +
{1 \ov p^2} Q(z) = u^2
}
with
\eqn\degFVl{
V(z) =  {f(r) \ov r^2}  \
}
and
\eqn\ghdrl{
Q(z) =  f \le({(d-1)^2 \ov 4 r^d} +\nu^2 -{1 \ov 4}\ri) \ .
}
When $z\rightarrow 0$ (i.e. $r \to \infty$) the potential \degFVl\ is constant and equal to $1$, which is
different from \bougd. As a result the WKB solution near the boundary
behaves like:
 \eqn\eSNk{
g_{\om p}^{(wkb)}\sim e^{\pm i p (u^2-1)^{1\ov 2} z}
 }
which is not the correct behavior \Bonh. This can be fixed as follows. One can solve the full equation \TeomD\ for $g(z)$ near $z \to 0$ and then match it with the WKB solution \eSNk.

For $u>1$, solving \TeomD\ near the boundary $z \to 0$ where the potential $V(z)\sim p^2+z^{-2}(\nu^2-{1\ov 4})$, we find that
 \eqn\bee{
g_{\om p}(z)\propto z^{1\ov 2} J_{\nu}(p (u^2-1)^{1\ov 2} z) \ .
 }
In the large $p$ limit, for any $z\neq 0$, we can expand \bee\ for large $pz$, giving
\eqn\turnzzero{
g_{\om p}(z)\sim \cos\left( p (u^2-1)^{1\ov 2} z-\pi \le({1\ov 4}+{\nu \ov 2}\ri)\right)}
which can be matched with the WKB solution and fixes  $g^{(wkb)}$ to be
\eqn\wkbpla{
g^{(wkb)}_{\om p}(z)\sim {1 \ov \sqrt{p(z)}} \cos\left(p W-{\pi\ov 4}(1+2\nu)\right),
 }
with
 \eqn\nfr{
 W=\int_0^z dz' p(z') , \qquad p(z) \equiv  \sqrt{u^2-f(r)r^{-2}}\ .
 }
Thus for $u > 1$, we get parallel result with \shde-\wdefB\ with $z=0$ as the turning point in place of $z_c$.  Also note there is an additional ${\pi \nu \ov 2}$ term
inside the argument for $\cos$ in \wkbpla\ which differs from \shde.

When $u<1$ there is now a turning point at $z>0$; apart from possible exponentially subdominant terms\foot{For $u$ with a small imaginary part these can be obtained by a more careful WKB analysis as outlined in Appendix A.} the result is exactly the same as \shde\ with $p(z)=\sqrt{u^2-f(r)r^{-2}}$ and $p$ in place of $\nu$ as the large parameter.

\subsec{A Bohr-Sommerfeld quantization formula for quasinormal frequencies}

The positions of the quasinormal modes in the large $\nu$ or $p$ limit can be obtained by generalizing the WKB analysis of the previous subsections for complex values of $u$. A very general but somewhat involved procedure is described in appendix A. Here we outline the main ideas and give the final results.

The quasinormal modes are zeroes of the coefficient $ f(\om,p)$ of the second term in \hort. In the WKB expansion, as is clear from equation \wekN, $f (\om,p)$ can have zeros
only if certain exponentially subdominant term becomes of the same order as the dominant term \wekN, and the two cancel each other. The
values of $u$ at which certain subdominant terms become of the same order as the leading
 term form dimension one lines in the complex-$\om$ plane, the so-called anti-Stokes lines of the expansion. The quasinormal modes are then special points on the anti-Stokes lines at which the two contributions exactly cancel each other.

For real $u$ the WKB approximation of the function $g_{\om p}(z)$ for large $z$ is given by \shde. To find quasinormal modes we first analytically continue \shde\ to lower complex plane. We start with large positive values of $u$ and after giving $u$ a small negative imaginary part, we can rewrite \shde\ as\foot{For a square root function like the one in $\kappa(z)$, we take the branch cut to be along the negative real axis.}
\eqn\shdef{
g_{\om p}^{(wkb)} =\lim_{\ep \rightarrow 0}{ \ep^{\nu} \ov  \sqrt{-i \kappa (z)}} \le(e^{\nu \int_{\ep}^{z}dz'\kappa(z')
 - i {\pi \ov 4} }+e^{\nu \int_{\ep}^{z_c}dz'\kappa(z') -\nu \int_{z_c}^z dz' \kappa(z')
 + i {\pi \ov 4}} \ri)}
where $\ep$ is an IR cutoff near the boundary and we have expressed the normalization factor $A$ in \shde\ explicitly using \zzdef. In the first term
the integral in the exponential of $A$ in \zzdef\ and that for $W$ in \shde\ combine into a single contour integral from $\ep$ to $z$, which does not depend on the turning point.
In contrast the second term depends explicitly on the turning point $z_c (u)$, since we have to evaluate two integrals one from $\ep$ to the turning point $z_c (u)$ and a second one from $z_c (u)$ to $z$. Near the horizon $z\rightarrow \infty$, the first term in \shdef\ behaves as $e^{i \nu u z}$ while the second as $e^{-i \nu u z}$, which can be identified respectively with the two terms in \hort. Note that had we given a positive imaginary part to $u$ the role of the two terms would be interchanged, i.e. it would be the term behaving as $e^{i \nu u z}$ that depends on the turning point $z_c (u)$.

As one varies $u$ in the lower half plane, there are special values $u_b$, at which $z_c (u)$ merges with some other turning point of equation \turnPw, which we denote as $z_T (u)$. More explicitly, at $u_b$, we have $z_c (u_b) = z_T (u_b)=z_b$ and $V'(z_b)=0$. In the case that $z_T (u)$ is an active turning point, it gives a subdominant contribution (except on the anti-Stokes line specified below) to $g_{\om p}^{wkb}$ of exactly the same form as the  second term \shdef\ with $z_c$ replaced by $z_T$. There is no correction to the first term in \shdef, since it does not dependent on the turning point. Thus in the large $z$ region,
the ratio of the contributions from $z_c$ and $z_T$ to $f(\om,p)$
in~\hort\  is  given by
\eqn\ratio{
e^{2 \nu \ZZ(z_c,z_T)} \equiv e^{2 \nu \int_{z_{c}}^{z_{T}} dz' \kappa(z')} \ .
}
The two contributions are of the same order when $\ZZ(z_c,z_T) \equiv  \int_{z_{c}}^{z_{T}}dz' \kappa(z')$ is pure imaginary. This condition determines an anti-Stokes line in the complex $u$ plane passing through $u_b$, when the ratio becomes unity. More careful WKB analysis shows that the anti-Stokes line is in fact a half line since along the other half line $z_T$ ceases to be an active turning point and needs not to be considered. Which half line to be chosen can often be easily determined on physical ground without detailed analysis. For real spatial momentum, the anti-Stokes line should always emanate from $u_b$ in the direction away from the real $u$-axis, since for real $u$ there is no exchange of dominance and thus the anti-Stokes line should not intersect the real $u$-axis. Note that $u_b$ is also a branch point for $z_c (u)$ and in our discussion above we have assumed that $z_c (u)$ is always the dominant turning point outside the anti-Stokes line, which is equivalent to  choosing its branch cut to be along the anti-Stokes line.

$f(\om,p)$ in \hort\ has zeroes when $e^{2 \nu \ZZ(z_c,z_T)} =- 1$, which leads to the condition
\eqn\rdbs{
2 \nu \ZZ(z_c,z_T) =2 \nu \int_{z_{c}}^{z_{T}}dz' \kappa(z')= \pi i (1+2n), \qquad n=0, 1, \cdots \ .
}
Note that only half of the integers are chosen in \rdbs, reflecting the half line nature of the anti-Stokes line. Equation \rdbs\ determines a line of quasinormal poles
which starts at $u_b$. To find all quasinormal poles we can repeat the procedure for all points where $z_c (u)$ merges with some other turning point.

To summarize, we now give the full procedure for finding the quasinormal modes
using \rdbs:

\item{1.} Find the turning point $z_c(u)$ from \turnPw\ for large positive $u$ and
analytically continue it to the lower half $u$-plane. Note that while \turnPw\ have multiple solutions $z_c (u)$ can be uniquely determined for a given large positive $u$. We will call $z_c (u)$ the physical turning point.

\item{2.} Find the roots $z_b$ to equation $\p_z V (z) =0$ and from the turning point equation \turnPw\ find the corresponding $u_b = V(z_b)$. At such a $u_b$
     two turning points of \turnPw\ merge together. However not all of them are relevant for finding quasinormal modes. One should {\it only} consider those $u_b$ while lie in the lower half plane and correspond to the merging of the physical turning $z_c (u)$ with some other turning point. We will call such a $u_b$, physical $u_b$.

\item{3.} For each physical $u_b$, there is a line of quasinormal poles determined by \rdbs, where $z_T (u)$ is the other turning point which merges with $z_c$ at $u_b$. The direction of the line should point away from real $u$-axis.

Equation \rdbs\ can be greatly simplified for the lower-lying quasinormal modes close to
$u_b$. Let $u=u_b+x $ with $x \sim O(\nu^{-1})$. Near
$z_b=z_T(u_b)=z_c(u_b)$  we can approximate the potential $V(z)$
as
$$
V (z) \approx  V(z_b) + \ha \p_z^2 V(z_b) (z-z_b)^2.
$$
We then find that
 \eqn\fssp{
z_{c,T} \approx z_b \pm  a, \qquad a \equiv 2\sqrt{u_b \, x \ov
\p_z^2 V (z_b)}
 }
and after integrating \rdbs\
 \eqn\valuax{
 \ZZ({z_c},z_T)
  \approx \pm {i \pi a^2 \ov 2} \sqrt{\p_z^2 V (z_b) \ov 2}
  = \pm { i \pi x \ov \delta}
 }
with
 \eqn\dksp{
 \delta \equiv \sqrt{\p_z^2 V \ov 2 V}\biggr|_{z_b} \ .
 }
Equation \rdbs\ then leads to the position $u_n$ of the quasinormal modes
 \eqn\dksP{
u_n=u_b+(n+\ha){\delta\ov \nu}, \qquad n=0,1,\cdots
 }
In terms of $\om=\nu u$, the modes are uniformly spaced near
$\om_b =\nu u_b$
 \eqn\usla{
\om_n = \om_b + (n+ \ha) \delta, \qquad n =0,1,\cdots
 }
with a spacing given by\foot{The formula below does not apply to the non generic case $\p_z^2 V(z_b)
=0$ which is realized if more than two turning points merge together or if $V(z_b)=0$ for finite $z_b$.} \dksp\ which is independent of $\nu$. The branch of the square root in \dksp\ should be chosen so that $\delta$ points away from the real axis at $u_b$.  Note that for modes which are of order $O(\nu)$ from $\om_b$ in
the $\om$-plane, \usla\ is no longer valid and one needs to use
\rdbs.

For large $p$ and $\omega$ but finite $\nu$ we can use the limit \defchi\ to obtain expressions for the quasinormal modes.
We will start from large real $u$ and analytically continue in the lower half plane.
As described earlier the boundary $z=0$ should be considered as the turning point for $u>1$. At $u=1$ this exchanges dominance with ($z_T$) present for $u<1$. An anti-Stokes line starts from $u=1$. On this line we find poles for:
\eqn\urosl{
2p \ZZ(0,z_T) =i\pi(1+\nu+2 n), \qquad n=0,1,\cdots
}
where the difference with respect \rdbs\ comes from the extra ${\pi \nu \ov 2}$ in \wkbpla\ with respect to \shde.

\newsec{Applications to a CFT on $\IR^{1,d-1}$}

We now apply the results of previous section to
find the quasinormal frequencies of scalar perturbations of
an AdS black brane with metric \bhmetric, which correspond to
poles of retarded Green functions for a CFT on $\IR^{1,d-1}$ at finite temperature.

Let us first consider $k=0$, in which case
\eqn\ghy{V(z) = f(r(z)) = r^2 - {1 \ov r^{d-2}} \ .}
The turning point equation \turnPw\ has coincident solutions for those values of $r$ such that $V'(r)=0$ that is
 \eqn\siaE{
\eqalign{2r+{d-2\ov r^{d-1}}=0 \ . \cr
 }}
Among $d$ solutions of \siaE\ only the following correspond to physical ones, i.e.
where the physical turning point $z_c (u)$ merges with some other turning point,
 $$
r = \le({d-2\ov 2}\ri)^{1\ov d}e^{\pm i {\pi \ov d}}
$$
The corresponding line of quasinormal modes starts at \eqn\braLo{
u_b = \pm c_d e^{\mp i {\pi \ov d}}, \qquad c_d = \le({d-2\ov
2}\ri)^{1\ov d} \sqrt{d \ov d-2} \ .
 }
The spacing of modes near $u_b$ can also be easily computed from \dksp\
 \eqn\spanS{
 \delta = \pm \sqrt{d} c_d e^{\mp i {\pi \ov d}},
 }
where each sign of \spanS\ should be paired with that of \braLo.
Expanding \rdbs\  for large $u$  we can find the mode spacing as $|\om|\rightarrow \infty$:
\eqn\deltinf{\delta = \pm  d \, \sin {\pi \ov d} \, e^{\mp i {\pi \ov d}}}

 Except for $d=4$ (i.e. AdS$_5$),
\spanS\ is different from the spacings \deltinf\ at large $\om$
although they do point to the same directions. This is expected as for $k=0$ since from \ghy\ that the phase of $V(z)$ is constant for $arg(z)=\pm{\pi\ov d}$ and therefore the mode line is straight. As we will see this changes for $k\neq 0$.

The explicit analytic expressions for the branch points and spacings of the quasinormal modes for $k \neq 0$ are rather
complicated since they involve roots of high order polynomial equations. We will only point out some important
features for $k$ large.

For a real $k$, the location of the quasinormal modes is similar to that for
$k=0$. There are four lines of modes. In particular, as $k^2 \to
+ \infty$, the branch points approach the real axis
 \eqn\nehab{
 u_b \approx \pm \le( k + \ha \le({d \ov 2}\ri)^{2 \ov d+2} e^{\mp
 {2 \pi i \ov d+2}} \, k^{-{d-2 \ov d+2}} + \cdots \ri)
 }
and the mode spacings near the branch points are
 \eqn\pilL{
 \delta \approx \pm  \sqrt{d+2} \le({d \ov 2}\ri)^{2 \ov d+2} e^{\mp
 {2 \pi i \ov d+2}} \, k^{-{d-2 \ov d+2}} + \cdots
 }
One can check that the angle between the lines of quasinormal modes and the real $u$ axis increases monotonically
as $k^2$ increases.
At large $\om$ the effect of a finite $k$ can be neglected and the mode spacing is again given by \deltinf\ therefore the mode lines are no longer straight.

\ifig\elpol{Numerical determination of one of the quasinormal mode lines in the complex $u$ plane for $k=20$, the straight lines tangent at $\infty$ and at the starting point are obtained from \deltinf\ and \pilL\ respectively.}{\epsfxsize=4cm \epsfbox{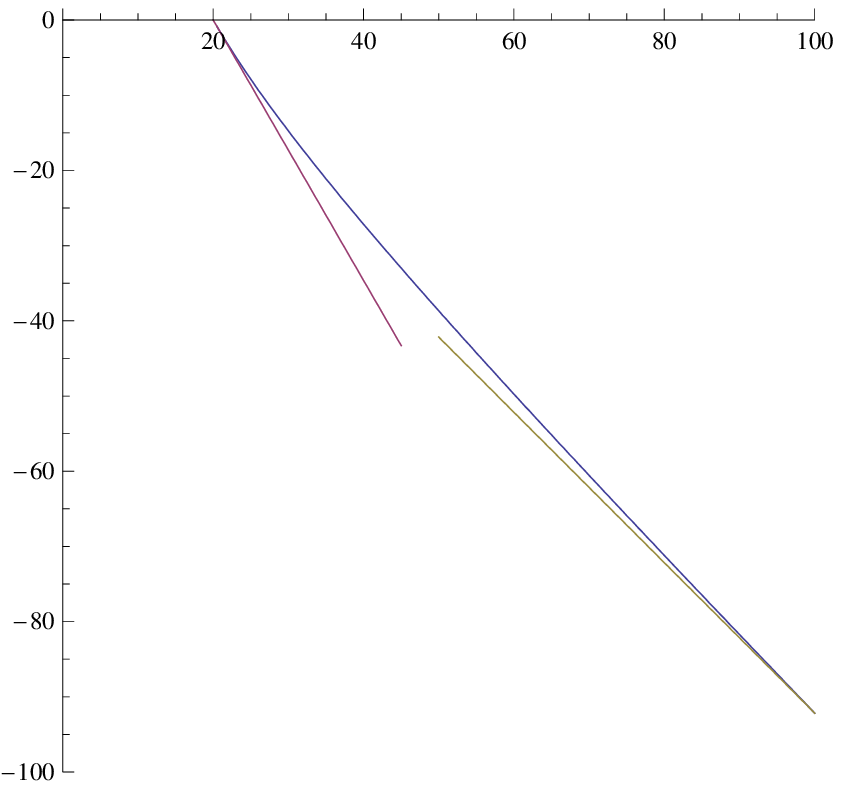}}

In \elpol, we plot a numerical determination of a mode line for $k=20$ in $d=4$ and the value of \rdbs\ along this line as a function of the real part of $u$.

At finite $\nu$, we can use the limit \defchi\ to obtain expressions for the quasinormal modes at large $p$ and $\omega$. As described there in this case $z=0$ (the boundary) should be considered as a turning point. The second turning point is
$$
r_T=(1-u^2)^{2\over 2-d}
$$
and merges with $z=0$ for $u=1$.
Equation \urosl\ then gives the lower-lying
quasinormal modes
 \eqn\rnfe{
\om=\pm p \pm  K \, e^{\mp i {2\pi \over d+2}} \, p^{2-d \over d+2} \, ( 1+\nu + 2n)^{2 d \over d+2},\;\;\;\; n=0,1, \cdots
 }
with $K$ a numerical factor given by
$$
 K^{d+2\ov 2d}={\sqrt{\pi} \Gamma[3/2+1/d] \ov \Gamma[1/d]}  \ .
 $$
Note that in~\rnfe\ the spacing between various modes increases with $n$ for $d > 2$.
In the limit $\om \rightarrow \infty$, one finds from \urosl\ that the spacing between quasinormal modes tends to the limit
$$\delta =  \pm d \sin\left({\pi \ov d}\right)e^{\mp i {\pi\over d}} $$ which agrees with formula \deltinf.

Note that in deriving \nehab\ and \pilL, we first take the large $\nu$ limit and then $k$
large, while \rnfe\ was obtained by taking $p \to \infty$ with $\nu$ fixed.
Taking $n=0$ and extrapolating \rnfe\ to the regime $p= k \nu$ with $\nu$ large, we find it  gives an expression which agrees with \nehab\ (after multiplying \nehab\ by $\nu$) up to the factor of $K$. Thus we see the two limits do not quite commute.

We now summarize our results by writing down various expression for $d=4$ explicitly.
We will only write down the sequence on the lower right quadrant. The sequence
on lower left quadrant can be obtained by reflection with respect to the imaginary $\om$-axis. From \braLo--\deltinf, in the large $\nu$ limit with $k=0$ (or small $k \equiv {p \ov \nu} \ll 1$),
 \eqn\eokr{
    \om_n \approx (1 - i) \pi T (\nu + 2 n + 1) ,  \qquad n=0,1,2,\cdots
    }
 where we have restored the temperature dependence explicitly by using $T = {1 \ov \pi}$.
 \eokr\ differs from expressions in the literature~\refs{\ricar} valid at large $\om$ which can be obtained by replacing $n\rightarrow n+ {i\ov 2}\log(2)$. This is no surprise as the approximation used in~\refs{\ricar} is not valid for $\nu \sim \om\rightarrow \infty$. The approximation of \refs{\siopsis} which uses a combination of large $\nu$ and $k=0$ also results in \eokr. We expect \eokr\ to be reliable for large values of $\om$ as the regime of validity of~\refs{\ricar} is $\om\gg \nu^{2-{2\over d}}$.

 In the limit of large $\nu$ and $k \equiv {p \ov \nu} \gg 1$, \nehab\ and \pilL\ give the lower-lying quasinormal frequencies
 \eqn\rnme{
 \om \approx p +  2^{-{2 \ov 3}} e^{-{i \pi \ov 3}} (\pi T) \le({p \ov \nu \pi T} \ri)^{-{1 \ov 3}}
  \le(\nu +  \sqrt{6} \le(2n +1 \ri) \ri), \quad n=0,1,\cdots
 }
In the limit $p \gg \nu$ with $\nu$ fixed, we find from
 \eqn\rnfNe{
\om= p +  K \, e^{-  {i\pi \over 3}} \,(\pi T)^{4 \ov 3} p^{-{1 \ov 3}} \, ( 1+\nu + 2n)^{4 \over 3},\;\;\;\; n=0,1, \cdots
 }
with $K = 0.344$. Note that \rnfNe\ also applies to massless fields in the bulk which correspond to $\nu = 2$.

For $p \gg (\De-1)^4 \pi T$, the poles in~\rnfNe\
 have an imaginary part which is much smaller than the temperature scale. However, in~\rnfNe\ the imaginary parts are  of the same order as the spacings between the poles. Thus in the spectral function we expect not to see individual peaks for each pole, rather a broad peak close to the light cone $\om = p$ reflecting the effects of a large number of poles. Note that the limit in going to zero temperature is not continuous, since for a CFT on $\IR^3$ at finite temperature, the temperature sets the scale for the system.
At zero temperature the pole line  \rnfNe\ is replaced by a branch cut
starting at $\om=p$.

The emergence of the time scale $p^{1/3}$ as implied by the imaginary part in~\rnfNe\
can also be seen by looking at time-like geodesics propagating in the bulk.
For a time like geodesic we have
\eqn\rkrk{
{d t\over dr}={E\over f(r)\sqrt{E^2-f(r)(1+p^2/r^2)}}, \qquad
f (r) = r^2 - {1 \ov r^2}
 }
where $E$ and $p$ are the conserved quantities along the geodesics corresponding to momentum in $t$ and in the spatial directions. For large $p$, equation \rkrk\ has the following
scaling limit
 \eqn\rjrn{
E^2-p^2 \sim p^{2\over 3}, \qquad r \sim p^{1 \ov 3}, \qquad t \sim p^{1 \ov 3} \ .
 }
Thus such a geodesic approach the boundary more and more as $p$ is increased and the time
the geodesic spends in the region where $r$ is $O(p^{1\over 3})$ is also of order $p^{1\over 3}$. During this time the geodesic propagates close to the boundary light cone.

To conclude this section, let us briefly comment on the quasinormal modes for pure imaginary $k$'s. Imaginary $k$'s are of interest as they play a role for probing the
geometry inside the horizon and near the singularity using Yang-Mills theory~\refs{\FestucciaPI}. For $k^2 = -q^2 < 0$, there is an additional line of quasinormal modes
along the imaginary $u$ axis with branch points located at $u_0 =
- i E_c$. For $q^2 \ll 1$, one finds that
 \eqn\imajp{
 E_c \approx \sqrt{2 \ov d} \le({d \ov d-2}\ri)^{-{d-2 \ov 4}}
 q^{-{d-2 \ov 2}} \gg 1
 }
and the spacing between modes near the branch point is given by:
 \eqn\ueros{
 \delta = - {i } {d \ov \sqrt{2}} \le({d-2 \ov d} \ri)^{d
 \ov 2} q^{- (d-1)} \ .
 }
$E_c$ decrease monotonically to zero as
$q^2$ increases to $1$. For $q^2\sim 1$ the branch point reaches $u=0$ and the two merging turning points tend toward $z=+\infty$. The mode spacing can still be found using \dksp\ because $\lim_{u_0\rightarrow0}{\partial_z V(z)\over V(z)}|_{z_0}$ is well defined.
This results in a spacing $\delta$ between successive
modes $\delta=i d+O(q-1)$.

\ifig\wkbja{The potential \degFV\ for $-k^2=q^2 > 1$ admits bound
states. In the $z$ coordinate the boundary is at $z=0$ and the
horizon at $z=\infty$.} {\epsfxsize=4cm \epsfbox{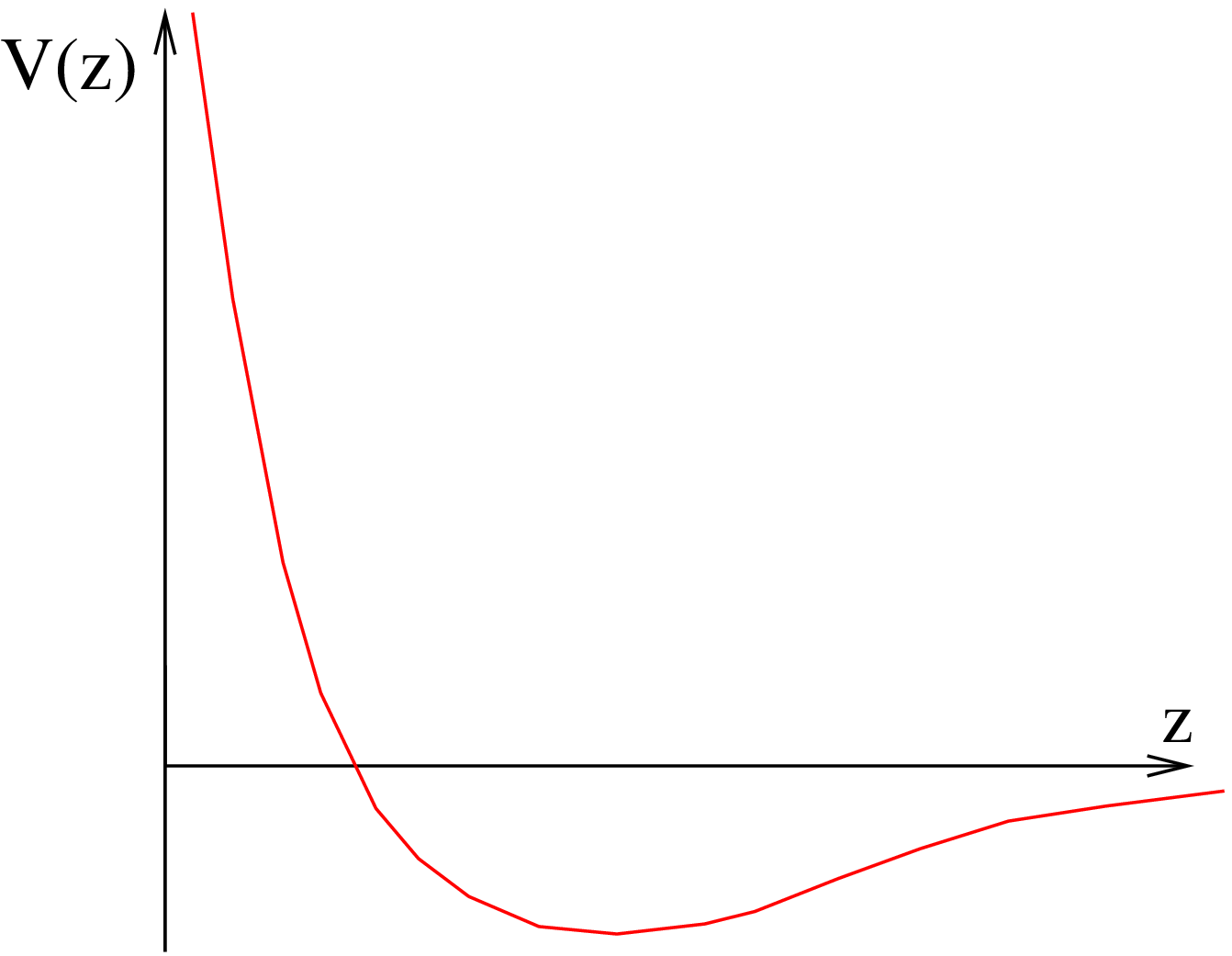}}

 \ifig\branchi{Schematic structure of the mode lines for (a): $k^2=0$, (b): $k^2>0$, (c): $-1 < k^2<0$,
(d): $k^2<-1$. At finite $\nu$, the modes are at discrete points along the lines.}{\epsfxsize=7.5cm \epsfbox{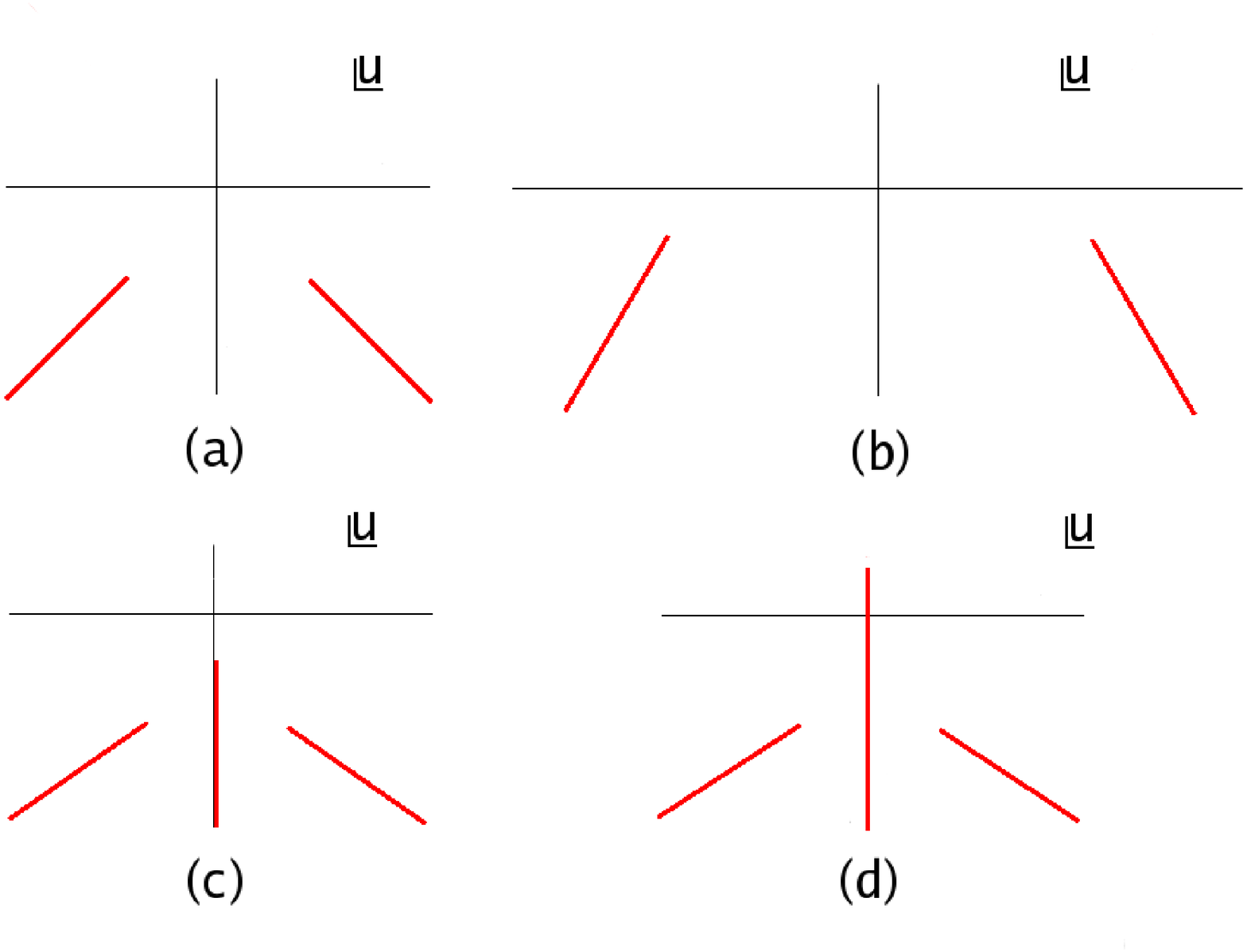}}

For $q^2> 1$, this line of modes along the imaginary $u$-axis crosses the real axis. This is due to the fact that for $q^2>1$, the Schrodinger problem \TeomD\ admits bound states, as is clear from the shape of the potential plotted in \wkbja. The modes lying in the
upper half plane are given by the bound states. In \branchi, we plot schematically the configurations of the quasinormal mode lines for various values of $k^2$.

\newsec{Long-lived quasiparticles for a CFT$_d$ on $S^{d-1}$}

\ifig\pot{The left figure is a schematic plot for the potential
$V(z)$ for $k > k_c$.  The resulting quasinormal mode lines are shown
in the diagram at the right. In this plot we only show
the right half of the complex $u$-plane. The modes in the left
half are obtained by reflection with respect to the imaginary
axis. } {\epsfxsize=7cm \epsfbox{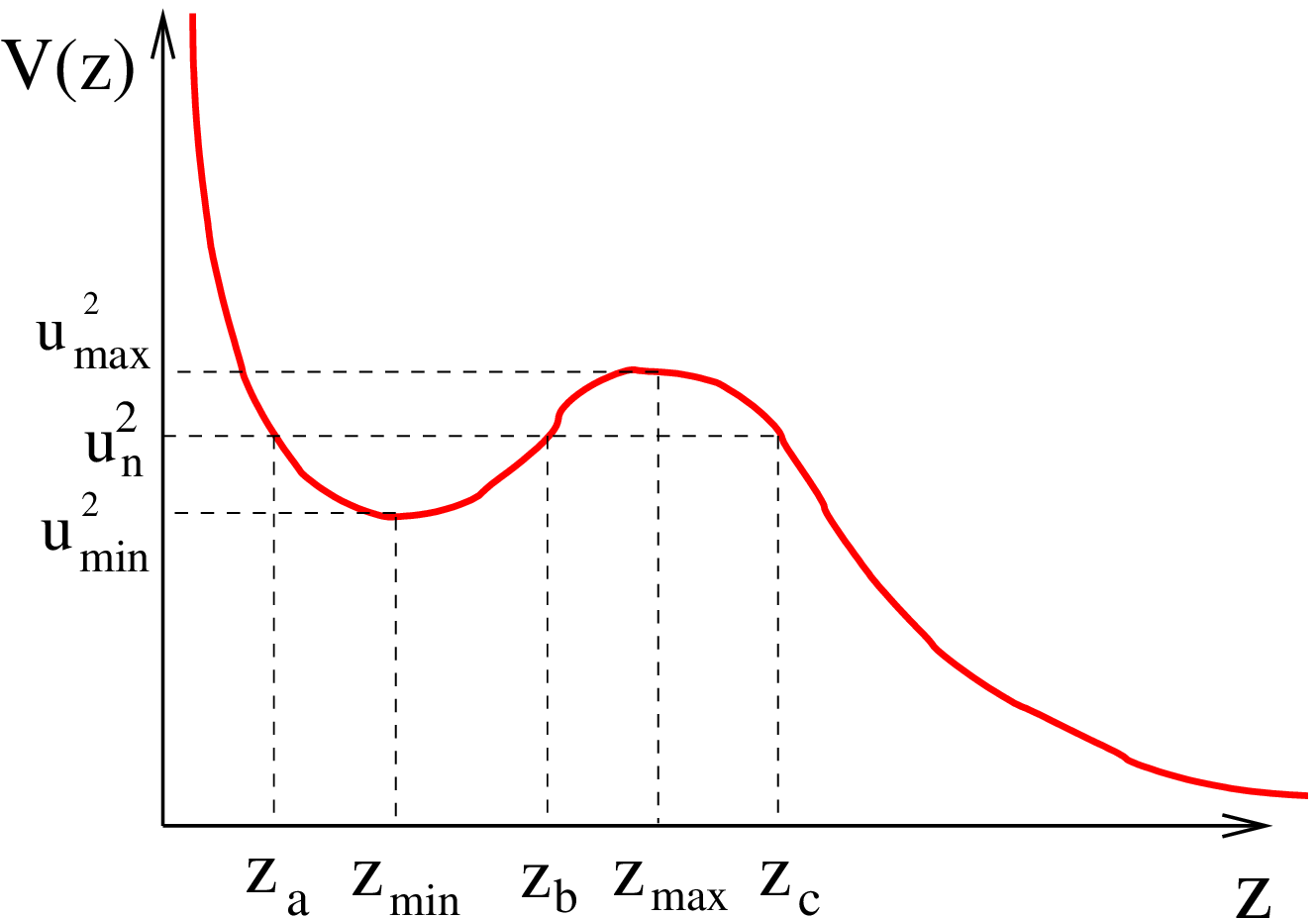}\epsfxsize=5cm
\epsfbox{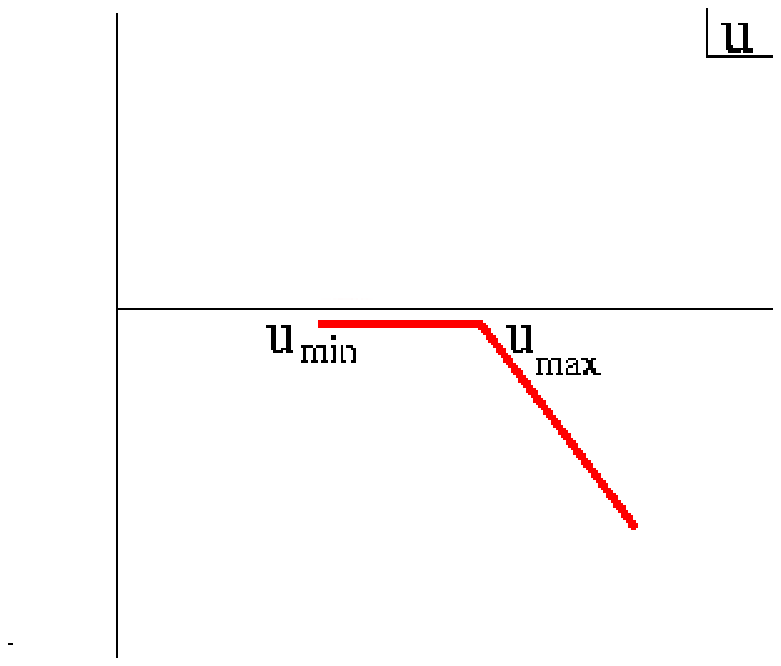}}

In this section we consider quasinormal modes for black hole metric \adsbh\ and \defF, which correspond to poles of retarded functions for a CFT on $S^{d-1}$ at finite temperature.
In this case the quasinormal modes are labeled by angular momentum $l$ on $S^{d-1}$
and the corresponding large $\nu$ limit is now
\eqn\defliml{\om=\nu u,\;\; (2l+d-2)=2\nu k,\;\;\nu\gg 1 \ .
}
For small values of $k  \in \IR$, the potential $V(z)$ \degFV\ is a monotonically decreasing function
along the real $z$ axis. The pattern of quasinormal modes is very similar to that of the black brane discussed in the last section and is qualitatively captured by \branchi.
Here we simply note that for $d=4$ at $k =0 $ (or $k \ll 1$) one gets from \rdbs\ the following simple formula\foot{For other dimensions, the expression is more complicated.  For $d=6$ and $d=3$  it is possible to evaluate \rdbs\ for $k=0$ in terms of elliptical integrals.} (for the sequence in the lower right plane)
\eqn\polgen{\eqalign{
\omega_n & = {2\pi\ov
\BB}\le(\nu+2 n + 1\ri)\qquad n=0,1,2,... \ }}
with $\BB$ given by
 \eqn\rrkN{
 \BB = {2 \pi \ov r_1 - i r_0} , \qquad {\rm with} \qquad {\rm Im} \, \BB = \beta
 }
where $r_0$ is radius of the horizon and $r_1$ is defined by\foot{Note the mass of the black $\mu$ can also be written as $\mu = r_0^2 r_1^2$.}
$$
r_1^2 = r_0^2 +1 \ .
$$
The other sequence of modes are obtained from the reflection of \polgen\ across the imaginary axis. \polgen\ is consistent with previous results~\refs{\ricar,\siopsis}
in the overlapping region of validity.

For larger
$k$ however there exists a critical value $k_c$ such that for $k>k_c$ the potential is no longer
monotonic and looks like the left plot of \pot. At $k=k_c$ we have $z_{min}=z_{max}$ and at $z= z_{min}=z_{max}$
both the first and second derivative of $V(z)$ are zero. By using the relation \tortoise\ between $z$ and $r$ and the explicit expression of $V(z)$ \degFV, $k_c$ can be found from the largest positive root of the
coupled equations $V'(z) =0$ and $V''(z) =0$.

For example, for $d=4$ 
one finds
 \eqn\kcew{
 (k_c^2-\mu)^3-27\mu^2
k_c^4=0  \ .
 }
For large $\mu$, \kcew\ gives\foot{which is pretty good already for $\mu > 2$, required for the black hole solution to dominate the thermal ensemble.} $k_c \approx 3 \sqrt{3} \mu$.
For generic dimensions $d > 2$, one finds that for large $\mu$,
 \eqn\ekns{
 k_c \sim \mu^{2 \ov d-2} \ .
  }

The form of the potential \pot\ for $k>k_c$ implies that it is
classically possible for a particle with sufficient angular

momentum to be in a bounded orbit outside the horizon of the black
hole as the centrifugal potential provides a barrier to its
falling in the horizon. Quantum mechanically however there will be
a nonzero probability for the particle to tunnel through the
centrifugal potential barrier and be absorbed by the black hole.

Applying \rdbs\ to the situation for $k > k_c$ (\pot) simply reduces to the standard Bohr-Sommerfeld quantization in the potential well (see also~\refs{\GrainDG} for the case of a massless perturbation in $d=3$). We thus find
resonances  whose energies are given by $\om_n = \nu u_n$ with
$u_n$ determined by
 \eqn\Djsj{
2 \nu \int_{z_a}^{z_b} dz
\sqrt{u_n^2-V(z)} =2\pi \le(n+\ha \ri), \qquad
n =0,1,\cdots
 }
where $u_{min}^2=V(z_{min})<u_n^2<V(z_{max})=u_{max}^2$ and $z_a,
z_b$ are real solutions to $V(z)=u_n^2$ (see \pot). The maximum
energy for these quasi-stable states is $u_{max}$. For small $n$ the first few energy levels will be close to $V(z_{min})$ and using \dksp\ we find that
 \eqn\posapp{
 u_n=u_{min} +
{1\ov \nu}\sqrt{\partial_z^2V(z_{min})\ov 2 V(z_{min})}\le(\ha+n \ri),
\qquad n =0,1,\cdots, \qquad u_{min}^2 = V(z_{min}) \ .
 }
These
quasi-stable states can tunnel through the potential barrier
between $z_b$ and $z_c$ (see \pot) to fall into the horizon, with
a decay rate given by
 \eqn\vndk{
\Gamma_n=\exp \le(-2\nu \int_{z_b}^{z_c}dz \sqrt{V(z)-u_n^2}
\ri) \ .
 }
Note that \vndk\ is not captured by~\rdbs\ since it is exponentially small in $\nu$.

Equations~\vndk\ and \posapp\ thus imply the existence of  poles with a very small imaginary part in boundary retarded Green functions. The quasinormal poles form lines in the complex $u$-plane
extending on the real axis from $u_{min}$ to $u_{max}$. At $u_{max}$ the lines start
deviating from the real axis as depicted in the right plot of \pot. For large $|u|\gg u_{max}$ the position pole line is qualitatively the same as the one for a monotonic potential.

The large $\nu$ limit is not essential to the existence of these resonances.
For large angular momentum and finite $\nu$ we can work in the limit
\eqn\dflimll{
\om \equiv p u,\qquad (2l+d-2) \equiv 2 p,\qquad p\gg 1
}
for which the WKB potential \degFVl\ is
 \eqn\ejrM{
 V (z) = 1 + {1 \ov r^2} - {\mu \ov r^d}
 }

 \ifig\Npot{Plot for the potential \ejrM\ with the turning points $z_b$ and $z_c$.} {\epsfxsize=6cm \epsfbox{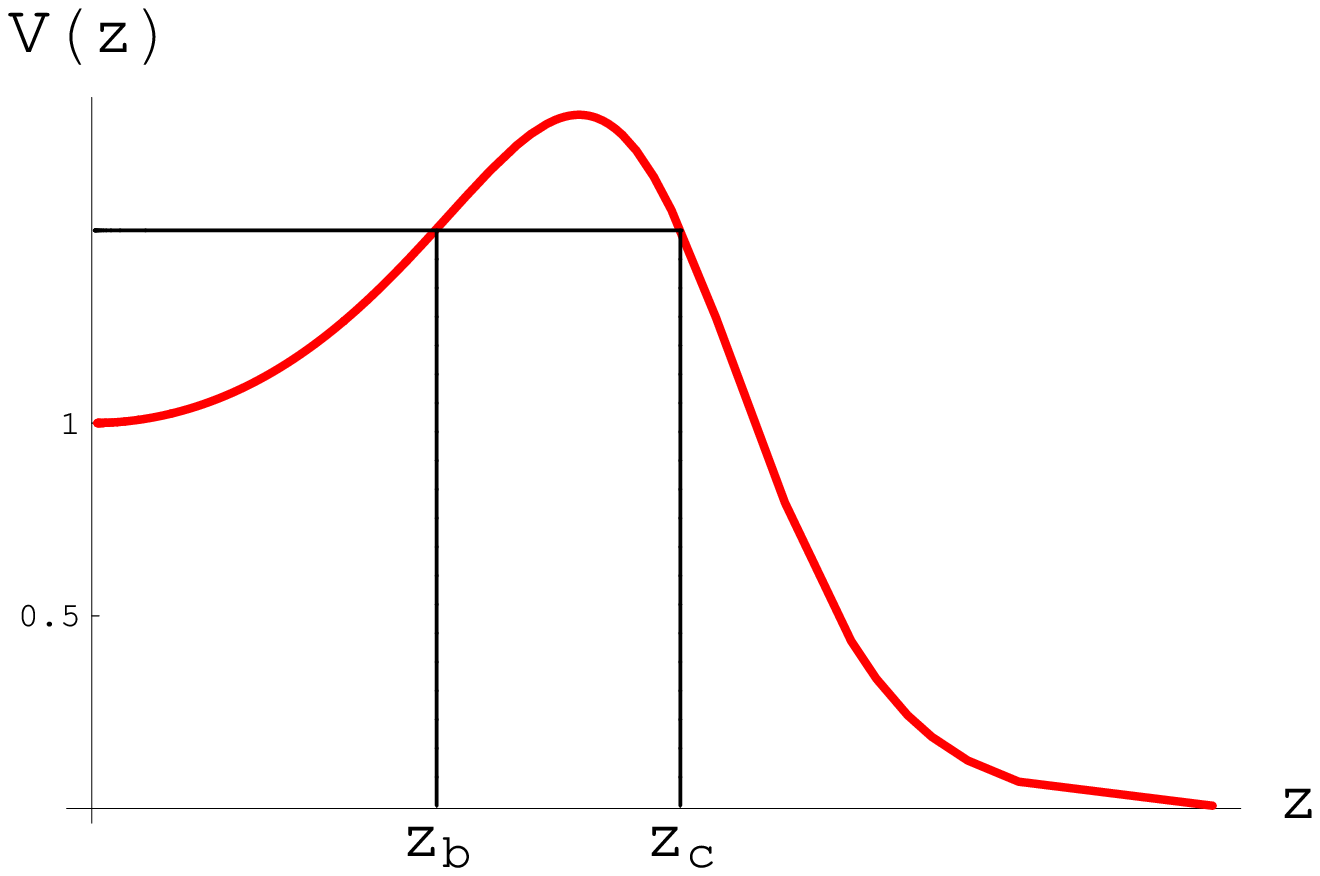}}

$V(z)$ has a maximum at (see \Npot)
 \eqn\rmmd{
 r_{max} = \le({d \mu \ov 2} \ri)^{1 \ov d-2}, \quad {\rm with} \quad
V(r_{max}) =  \le(1+{d-2\ov d \, r_{max}^2 } \ri)
 }
For a given $1< u < V (r_{max})$, there are long lived quasiparticles for
 \eqn\Djsjl{
2 p \int_{0}^{z_b} dz
\sqrt{u_n^2-V(z)} =\pi(2 n+1+\nu), \qquad
n \in 0,1,\cdots
 }
 with a decay rate
\eqn\vndkl{
\Gamma_n=\exp \le(-2p \int_{z_b}^{z_c}dz \sqrt{V(z)-u_n^2}
\ri) \
 }
where $z_b$ and $z_c$ are the turning points before and after the maximum as labeled in \Npot.
These long-lived quasiparticles extend from $u=1$ to $u^2=V(z_{max})$.
For $u\sim 1$ \eqn\posappl{
 \om_n =
p+(2n+1+\nu)\sqrt{\partial_z^2V(0)\ov 2 V(0)}=l+2n+ \De\ , \qquad n=0,1,\cdots \ .  }
Note that for equation \posappl\ to apply, we need $p (\sqrt{V (r_{max})} - 1) \gg  \Delta$. For $r_{max}$ large~(i.e. $\mu$ large), this condition can be written explicitly as
 \eqn\prnw{
 l \gg \De \, \mu^{2 \ov d-2} \sim \De \, T^{2d \ov d-2}
 }
where we have used that $\mu \sim T^{1 \ov d}$ for large $\mu$. Note that \prnw\ has the same scaling with $\mu$ as \ekns.

Notice that \posappl\ is exactly the spectrum of a scalar conformal operator of dimension $\De$ with angular momentum $l$ at zero temperature. That in the limit $l \to \infty$ we recover the zero temperature spectrum is not surprising. We expect states of energies in~\posappl\ not to be excited in the thermal ensemble due to the exponential suppression in $e^{-\beta l} $ and that the degeneracy of these states only increases with $l$ as a power.
However, it is rather curious the power in $T$ in the onset condition \prnw\ and \ekns\ for the appearance of  long-lived quasi-particles is much larger than $1$, as one would naively expect. For example for $d=4$, \prnw\ gives $l \gg \De T^4$. Another interesting feature is that for a given $l$ satisfying \prnw, the zero temperature spectrum \posappl\ only persists for a finite range of frequencies from $\om_0 = l + \De$ to $\om_{max} \sim l (1 + \mu^{-{2 \ov d-2}} )$. When $\om > \om_{max}$, finite temperature effects again set in.
It would be interesting to understand these issues better and to investigate whether there are some physical processes in which such long-lived quasi-particles play a dominant role.

To conclude this section, let us consider what happens to these
resonances in the limit
\scalU, which describes the boundary theory on $\IR^{d-1}$. In
momentum space, the limit \scalU\ can be described as
 \eqn\ehnas{
  {\om \ov  T} = {\rm finite}, \qquad {l\ov T} =  {\rm finite}, \qquad T \to
  \infty \ .
 }
In the limit $T \to \infty$, from equations \prnw\ and \ekns, the onset value of $l$ for appearance of long-lived quasi-particles scales with $T$ as $T^{2 d\ov d-2}$. It then follows that the frequencies
 and angular momenta of the resonances scale with $T$ at least as fast as $T^{2d
\ov d-2}>T$, which is much faster than \ehnas. Thus we conclude that these resonances  disappear in the limit \scalU. Indeed, as we discussed earlier, the potential
$V(z)$ for
\newf\ is always monotonic and there is no $k_c$.

\bigskip
\noindent{\bf Acknowledgments}

We would like to think N. Iqbal, H.~Meyer, K.~Rajagopal, T.~Senthil, and S.~Shenker for useful conversations.
This research is supported in part by
the Offices of Nuclear and High Energy Physics of the Office of
Science of
the DOE
U.S.~Department of Energy
under
contracts \#DF-FC02-94ER40818 and \#DE-FG02-04ER41286.
HL is also supported
in part by the A.~P.~Sloan Foundation and the U.S. Department
of Energy OJI program.

\appendix{A}{WKB analysis of the quasi-normal modes}

In this section we describe how to extend the WKB approximation to account for subdominant contributions to $g_{\om p}(z)$ which are responsible for the appearance of the quasinormal frequencies. We start by reviewing general techniques which are applicable to any WKB computation\refs{\BerryMount,\KnollSchaeffer}. We then apply them to our specific case pinpointing some general features of the WKB result stemming from the structure of the potential \potenP\ or \poeV. Finally we work out a couple of specific examples.

\subsec{Review of complex WKB}

By performing the rescaling \largnu\ equation \TeomD\ in the large
$\nu$ limit reduces to:
\eqn\eqwkb{
\le({1\ov \nu^2}{d^2\ov dz^2}-V(z)+u^2\ri)\psi(z)=0
}

we want to find the form of $g(z)$ defined in \Bonh\ for
$z\rightarrow \infty$. From there comparing with \hort\ we will obtain approximate expressions for the quasinormal frequencies.

Define $\ZZ(z_0,z)=\int_{z_0}^z dz \kappa(z,u)$ where
$\kappa(z,u)=\sqrt{V(z)-u^2}$

\ifig\turng{Pattern of Stokes lines near a turning
point}{\epsfxsize=5cm \epsfbox{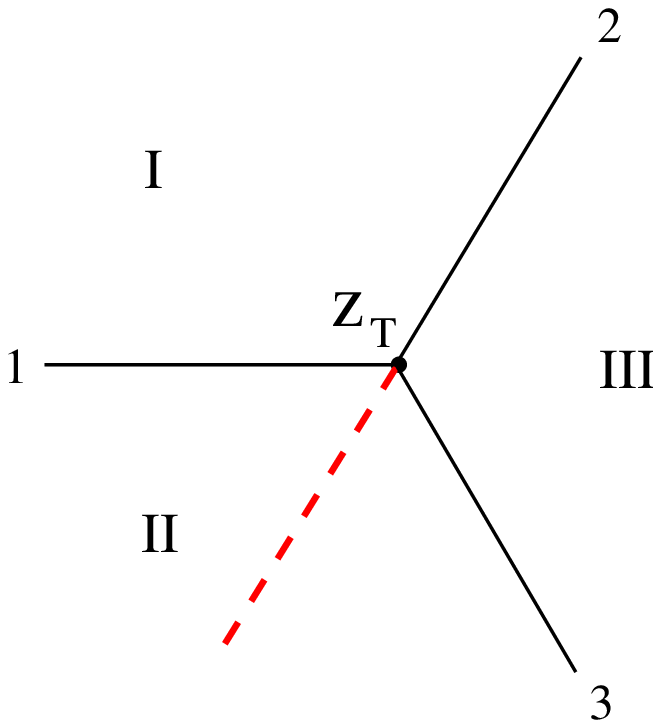}}

Close to a generic point $z_0$ the lines of constant imaginary
part for $\ZZ(z_0,z)$ do not intersect. However whenever $z_0=z_T$ with
$V(z_T)=u^2$ there are three of these lines\foot{More than three lines is not a generic situation and is not considered} converging towards
$z_T$ at a ${2 \pi\ov 3}$ angle; these are called Stokes lines. On
each of these $\ZZ(z_T,z)\in R$ and so the exponential factor $\kappa(z,u)^{-\ha} e^{\pm \nu \ZZ(z_T,z)}$ in
the WKB approximation to the solution is real and either decreasing or increasing along the line. Both $\kappa(z,u)^{-\ha}$ and $\ZZ(z_T,z)$ are
multi-valued functions around $z_T$ and we will define them by
introducing a branch cut extending from $z_T$ in region $II$ in
\turng\ . Suppose now $\ZZ(z_T,z)<0$ along line 1; then in the WKB
approximation of the solution decreasing along $1$ only the term
$\kappa(z,u)^{-\ha} e^{ \nu \ZZ(z_T,z)}$ is present in regions $I$
and $II$. We will find the WKB expansion of this solution in
region $III$ by applying the principle of exponential dominance
stating that crossing a Stokes line the coefficient of the
dominant term in the expansion does not change. The term
$\kappa(z,u)^{-\ha} e^{ \nu \ZZ(z_T,z)}$ is the dominant one on
line $2$ and so its coefficient is continuous and doesn't change
while crossing it. However in region $III$ we can also have a
contribution proportional to $\kappa(z,u)^{-\ha} e^{- \nu
\ZZ(z_T,z)}$ which will be the dominant one along line $3$. In
order to find its precise coefficient we analytically continue the
expansion we have in region $II$ up to line $3$ but in doing so we
cross the branch cut and so we get $ -i \kappa(z,u)^{-\ha} e^{-
\nu \ZZ(z_T,z)}$. Then in region $III$ the asymptotic expansion
is:
$$
\psi(z) \sim \kappa(z,u)^{-\ha}(e^{\nu \ZZ(z_T,z)}-i e^{- \nu \ZZ(z_T,z)}).
$$
If instead the branch cut was in region $I$ we would have
obtained in analogous way:
$$
\psi(z) \sim \kappa(z,u)^{-\ha}(e^{\nu \ZZ(z_T,z)}+i e^{- \nu \ZZ(z_T,z)}).
$$
Given a turning point $z_0$ its Stokes lines cannot intersect
unless in another turning point which is not a generic situation
and is excluded in what follows \foot{these non generic cases are
the ones that divide the parameter space in regions with
topologically distinct patterns of Stokes lines}, and therefore
divide the complex $z$ plane into three regions. Two cases are
possible:

a) The origin and $+\infty$ are in the same region then $z_T$ is
called inactive.

b) They are in different regions and $z_T$ is an active turning
point.

For each active turning point imagine shading the regions in which
the origin and $+\infty$ are; the intersection of all the shaded
regions will be called active region in the following. Starting
from the region containing the origin we can order the regions
which are active by adjacency up to the one containing $+\infty$.

For every turning point there is a region which doesn't contain
neither $0$ or $+\infty$. It is therefore possible to choose the
branch cuts defining $\sqrt{V(z)-u^2}$ and $\ZZ$ in such a way
that they do not cross into the active region.

Then in the active region we can globally define two wave-forms
$\exp(\pm \nu \ZZ(z_0,z))$. For each wave-form the Stokes lines
where it is dominant or subdominant are fixed by the sequence of
active turning points as shown in figure.

\ifig\activ{schematic representation of the active region. The red
dashed lines are branch cuts while the arrows are in the direction
of decreasing real part for $\ZZ$}{\epsfxsize=7cm
\epsfbox{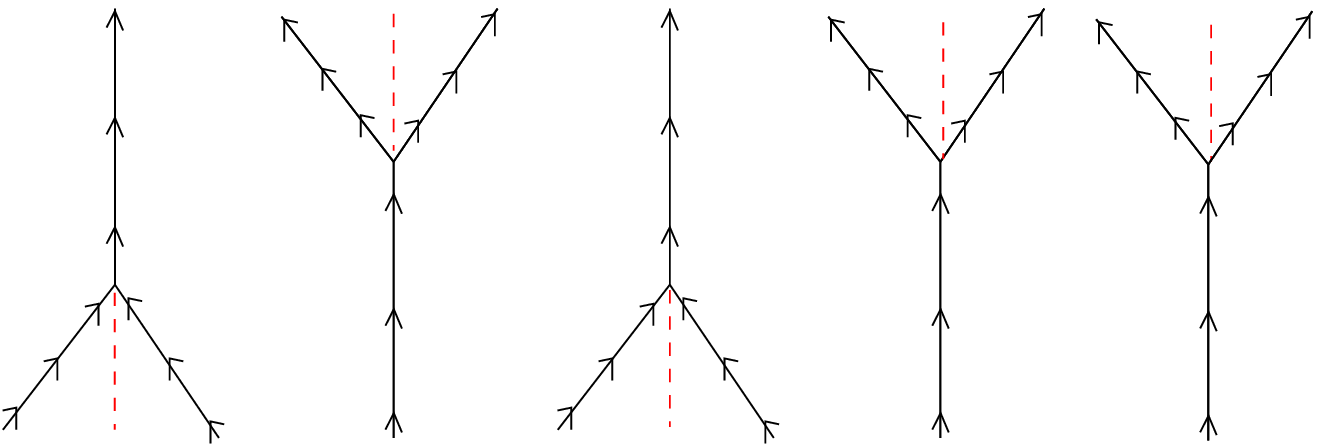}}

The WKB approximation to the desired solution is then given by
choosing the appropriate coefficients of the two wave-forms in
each region. When we cross one of the Stokes lines the coefficient
in front of the wave-form which is dominant along that line does
not change while the coefficient of the other wave changes
according to the appropriate connection formula. Two cases are
possible and are shown in figure (the arrows represent decreasing
$\ZZ(z_t,z)$):

{\epsfxsize=7cm \epsfbox{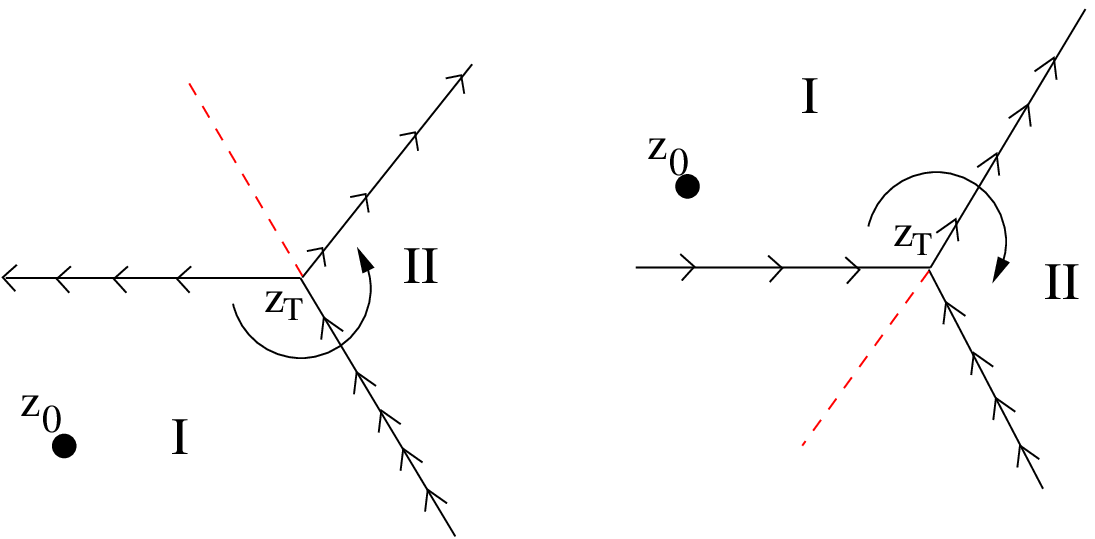}}

Suppose we start with the wave $A \exp(\nu \ZZ(z_0,z))+B \exp(-\nu
\ZZ(z_0,z))$ in region $I$ then in the first case the first term
is dominant while crossing the Stokes line and so its coefficient
doesn't change. By applying the connection formula we get in
region $II$:
$$
A \exp(\nu \ZZ(z_0,z))+B \exp(-\nu \ZZ(z_0,z))+ i A\exp(\nu
\ZZ(z_0,z_T)-\nu \ZZ(z_T,z))
$$
while in the second case the dominant term is the second one and
we obtain:
$$
A \exp(\nu \ZZ(z_0,z))-i B \exp(-\nu \ZZ(z_0,z_T)+\nu \ZZ(z_T,z))+
B\exp(-\nu \ZZ(z_0,z))
$$

\subsec{General remarks to the Black hole case}

The application of the method just described to the specific case
of the black hole background is in principle straightforward once
the pattern of stokes lines corresponding to the values of $k$ and
$u$ of interest is determined. Unfortunately the range of
possibilities is quite extended and we will content ourselves to
determine what are the turning points that give the dominant
contribution to the result in various regimes, and determine the
position of quasi-normal modes. In this subsection we will comment
on some general features of the WKB computation which arise from the exponential behavior of
$V(z)$ for $z\rightarrow \infty$. In particular this implies that the boundary $z=0$ and the horizon $z=\infty$
are separated by an infinite number of Stokes lines.

Being $V(z(r))$ a single valued function of $r$ it follows from the discussion of the
function $r(z)$ in Appendix B that $V(z)$ will have branch points at those points $z$ for which $r(z)=0$. In the
following we will use the determination for $r(z)$ described in Appendix B which has the following
properties:

\item{1} $Re(z)\rightarrow +\infty$ corresponds to $r(z)$ approaching the horizon.
$z=0$ corresponds to the boundary.
\item{2} The only branch points present for $Re(z)>0$ are located at
$z={\tilde \beta \ov 4}+i\; (n+\ha) {\beta \ov 2}$
where $n\in \IZ$ and ${1\over 4}(\tilde \beta +i \beta)$ is the location of the singularity in the complex $z$ plane.
For any branch point the branch cut extends on a line of
constant imaginary part for $z$ towards
$Re(z)=-\infty$.
\item{3} This determination is such that $V(z+i{\beta\ov 2})=V(z)$

\ifig\atr{pattern of Stokes lines for $Re(z)>0$ and $Re(u)>0,\;Im(u)<0$. The thick lines are the branch cuts extending from the singularity $S$ at $z={\tilb\ov 4}\pm {2n +1\ov 4}i \beta\;\;\;n\in \IZ$ while the boundary at
$z=0$ is denoted by $B$}{\epsfxsize=7cm \epsfbox{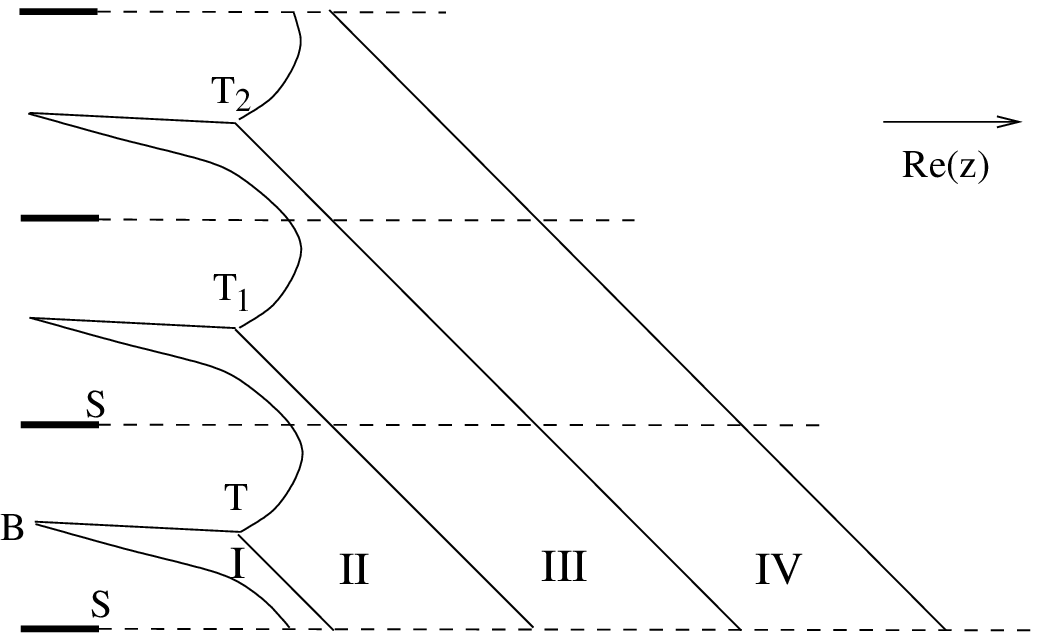}}

We will describe the consequences of the periodic structure of $V(z)$ with the simple case arising for $u$ close to the origin so that there is only one active turning point $T$ (and its periodic images), which is the one considered in \turnPw\ for $u^2\in R^+$. The pattern of Stokes lines is represented in \atr\ for $-{\pi\ov 2}<Arg(u)<0$.

Requiring $\ZZ$ to be continuous in the
active region fixes it up to a sign. In
the following we will conventionally
choose the determination of $\ZZ$ in
the active region in the following way:
For any two points
$z_1,\;z_2=z_1+i{\beta\ov 2}$ the
quantity $ \ZZ(z_1,z_{2})$ must be the
same due to the periodicity of the
function $V(z)$. By computing it for
$Re(z_1)\rightarrow +\infty$ we obtain:
$\ZZ(z_1,z_2)= \pm i
u\int_0^{i{\beta\ov 2}}dz$. We will
choose the sign of $\ZZ$ such that
$\ZZ(z_1,z_2)=-{\beta\ov 2} u$.

For $z$ in zone $I$ in figure, the WKB wave-function corresponding
to $g_{\om p}(z)$ defined in \Bonh\ is:
\eqn\firss{
\eqalign{&g_{\om ,p}(z) \sim A \kappa(z,u)^{-\ha} \exp(\nu \ZZ(z_T,z))\cr
&A=\lim_{\ep\rightarrow 0}\exp(\nu \ZZ(\ep,z_T))\ep^{\nu}=\exp(\nu
\ZZ^*(0,z_T))}
}

with $\ZZ^*(0,z)=\lim_{\ep\rightarrow 0}(\ZZ(\ep,z)+\nu
\log(\ep))$

Applying the rules given above we get for $z\in II,\;\;III,\;\;IV$.
$$
\eqalign{g_{\om p}(z)\sim &A\kappa(z,u)^{-\ha}\big( \exp(\nu \ZZ(z_T,z))+
i \exp(-\nu \ZZ(z_T,z))\big)\cr g_{\om p}(z)\sim &A
\kappa(z,u)^{-\ha}\big(\exp(\nu \ZZ(z_T,z))+ i  \exp(-\nu
\ZZ(z_T,z))[1+\exp(2\nu \ZZ(z_T,z_{T_1}))]\big)\cr g_{\om p}(z)\sim &A
\kappa(z,u)^{-\ha}\big(\exp(\nu \ZZ(z_T,z))+ i \exp(-\nu
\ZZ(z_T,z))[1+\exp(2\nu \ZZ(z_T,z_{T_1}))+\exp(2\nu
\ZZ(z_T,z_{T_2}))]\big)}
$$
For $z\rightarrow +\infty$ we obtain:
$$
g_{\om p}(z) \sim A \kappa(z,u)^{-\ha}\le(\exp(\nu \ZZ(z_T,z))+i
\exp(-\nu \ZZ(z_T,z))\left[ {1\ov 1-\exp(2\nu \ZZ(z_T,z_{T_1}))}\right]\ri)
$$
we can use the fact that as $z_{T_1}=z_T+i{\beta\ov 2}$ the
quantity $\exp(2\nu \ZZ(z_T,z_{T_1}))=e^{-\nu u \beta}$ to get
\eqn\gwkbexp{g_{\om p}(z)\sim A \kappa(z,u)^{-\ha}\le(\exp(\nu
\ZZ(z_T,z))+ i { e^{\nu u \beta/2}\ov 2\sinh( \nu u
\beta/2)}\exp(-\nu \ZZ(z_T,z))\ri) }

The factor ${ e^{\nu u \beta/2}\ov 2 \sinh( \nu u
\beta/2)}$ which is required by the analysis in Appendix B arises naturally due to the periodic nature of the potential.

For $z\rightarrow +\infty$ we have $\ZZ(z_T,z)\sim i uz$ and the following holds\foot{It would seem that this expression is
inconsistent with the general form of $g_{\om p}(z)$ described in Appendix B as the coefficients in front of $e^{i\nu u z}$ and $e^{-i\nu u z}$ are not related by $u\rightarrow -u$. However the form of the $WKB$ approximation to $g_{\om p}(z)$ does not have to be continuous in $u$. In fact \gwkbexp\ was found for $Im(u)<0$ while for $Im(u)>0$ the following expansion is valid
\eqn\gwkbexp{g_{\om p}(z)\sim A \kappa(z,u)^{-\ha}\le(\exp(-\nu
\ZZ(z_T,z))- i { e^{-\nu u \beta/2}\ov 2\sinh( \nu u
\beta/2)}\exp(\nu \ZZ(z_T,z))\ri). }}:

\eqn\Gsol{ g_{\om p}(z)\rightarrow {A\ov
i\nu u}\le( e^{i \nu u z+\nu \gamma}  + i { e^{\nu u \beta/2}\ov 2\sinh( \nu u
\beta/2)} e^{-i \nu u z-\nu \gamma} \ri).}
where \eqn\defgamma{\gamma=\lim_{z\rightarrow +\infty}(
\ZZ(z_T,z)-i\nu u z )}

Following our general discussion the quasinormal modes for $Im(u)<0$ show as zeroes in the
coefficient multiplying $e^{-i \nu u z}$ in \Gsol . We conclude that there are no quasinormal
 modes for small complex $\om=\nu u$.

 Due to the periodic nature of the potential however even in this simple case an infinite number of subdominant contributions to the WKB expression have to be considered for $z\rightarrow \infty$. This same complication will arise when other turning points enter the active region.

\subsec{The $k=0$ case for large $u$}

Increasing $u$ the pattern of Stokes lines changes as a second turning point becomes active. The situation is
exemplified by the figures below representing the Polya vector field~\refs{\PolyaLatta} for the function\foot{This is just the same as the Polya vector field for $\kappa(z,u)$ represented on the $r(z)$ complex plane. The $r$ coordinate is simpler to use in numerical applications} ${\kappa(r,u) f(r)^{-1}}$ for $\mu=4,\;\;d=8$ and various values of $u$. The singularity is represented by a purple dot at $r=0$ while the horizon is a green dot.
In the first figure which is representative of the situation for small $u$ there is only one active turning point represented in red. As the norm of $u$ is increased a second turning point (in blue) enters in the active region (figure 2). For $u$ close to the real axis the first turning point is the dominant one (figure 3), however as the phase of $u$ becomes more negative the two turning point exchange dominance (figure 4) and finally it is the second turning point which is dominant for even more negative phase of $u$ (figure 5).
For $|u|\rightarrow \infty$ the red turning point which we will denote as $T$ approaches the boundary $z_T \rightarrow 0$ while the blue turning point ($K$) approaches the singularity $z_K\rightarrow {1\ov 4}(\tilb+i \beta)$.
\ifig\plots{Polya plot of ${\kappa(r,u) f(r)^{-1}}$ for $\mu=4,\;\;d=8$ and $u=1.7e^{-i{\pi\over 16}}$, $u=1.82e^{-i{\pi\ov 16}}$ and $u=2e^{-i{\pi\ov 16}}$, the lines in color are sketched for visual purposes and do not represent a faithful integration of the field}{\epsfxsize=3.5cm \epsfbox{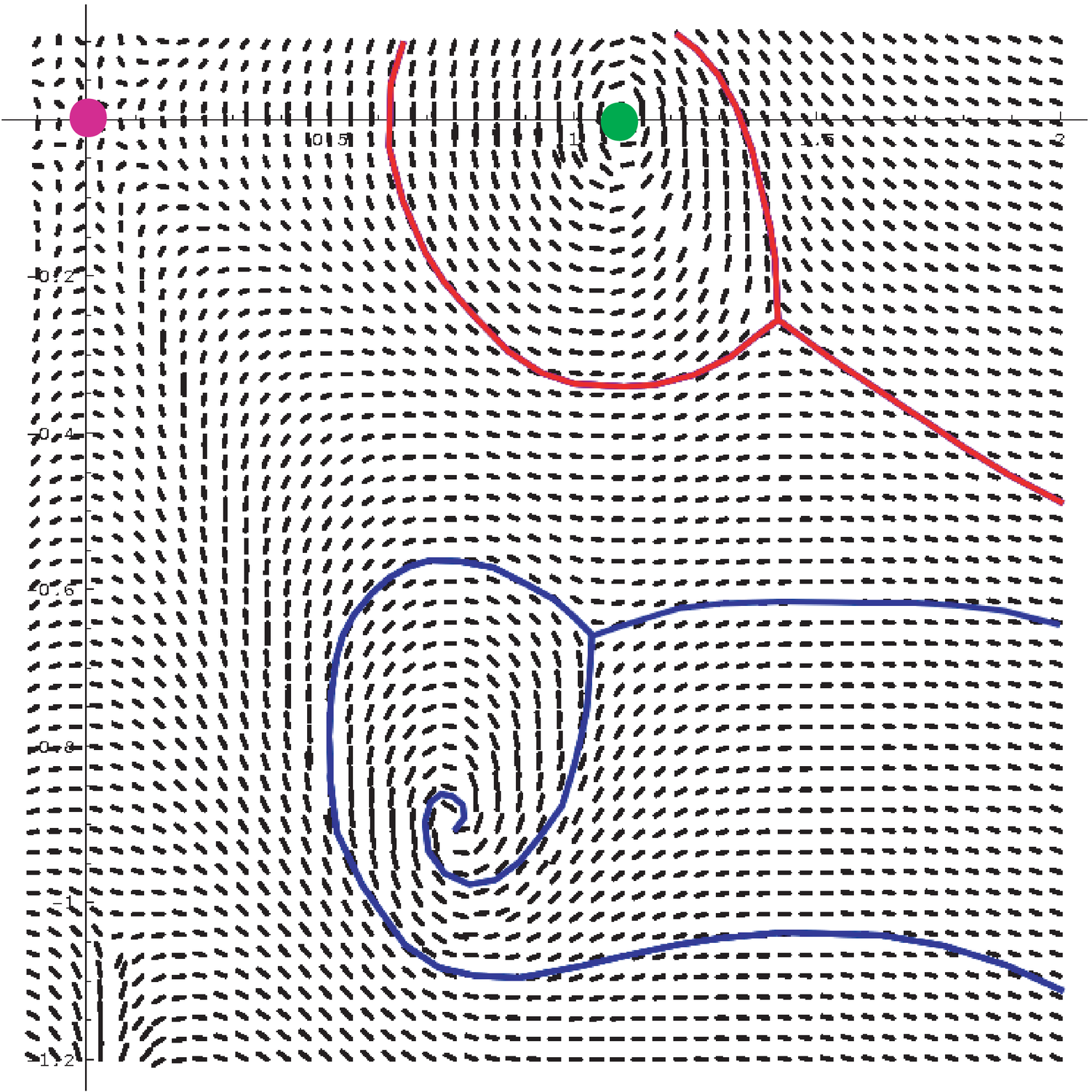} \epsfxsize=3.5cm
\epsfbox{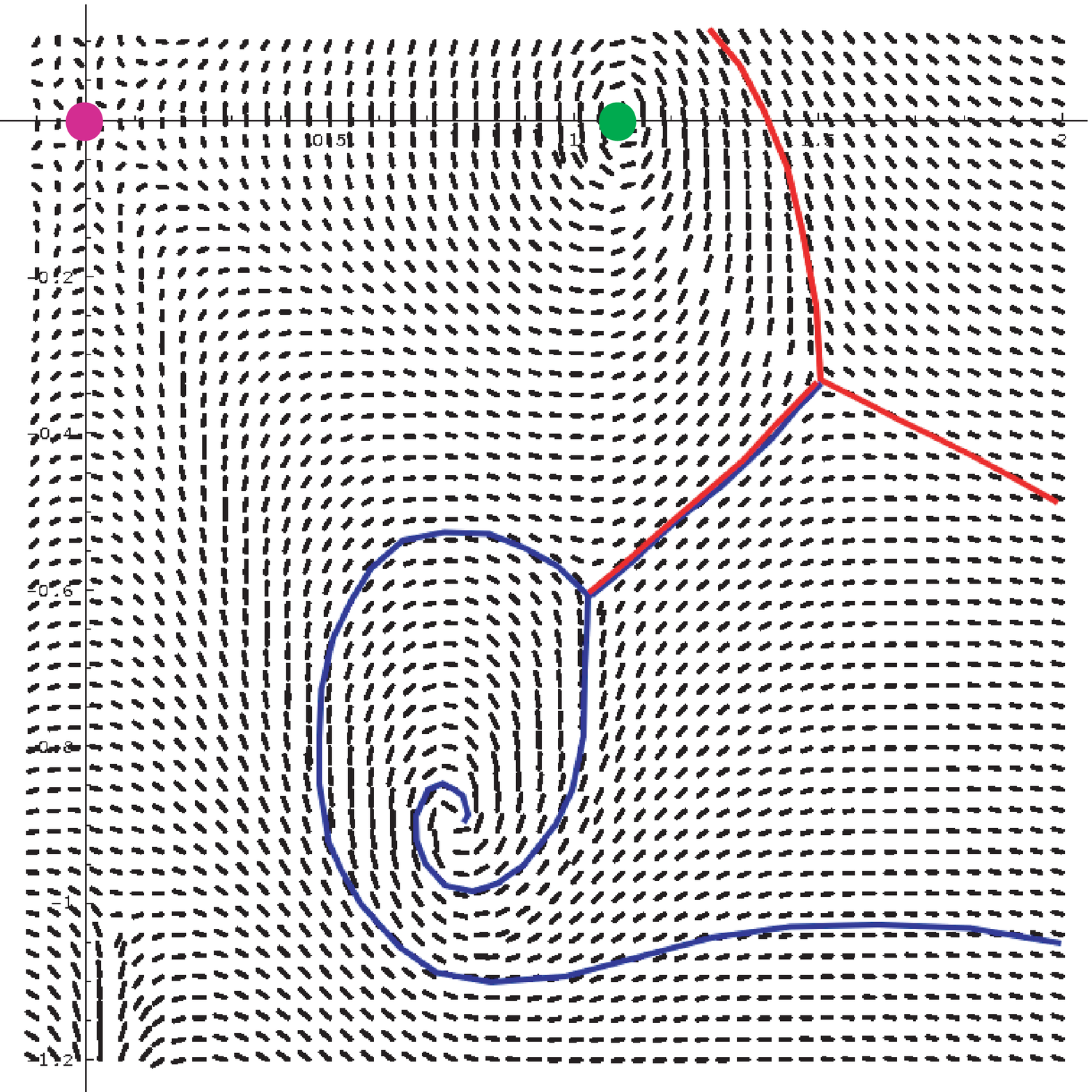}\epsfxsize=3.5cm \epsfbox{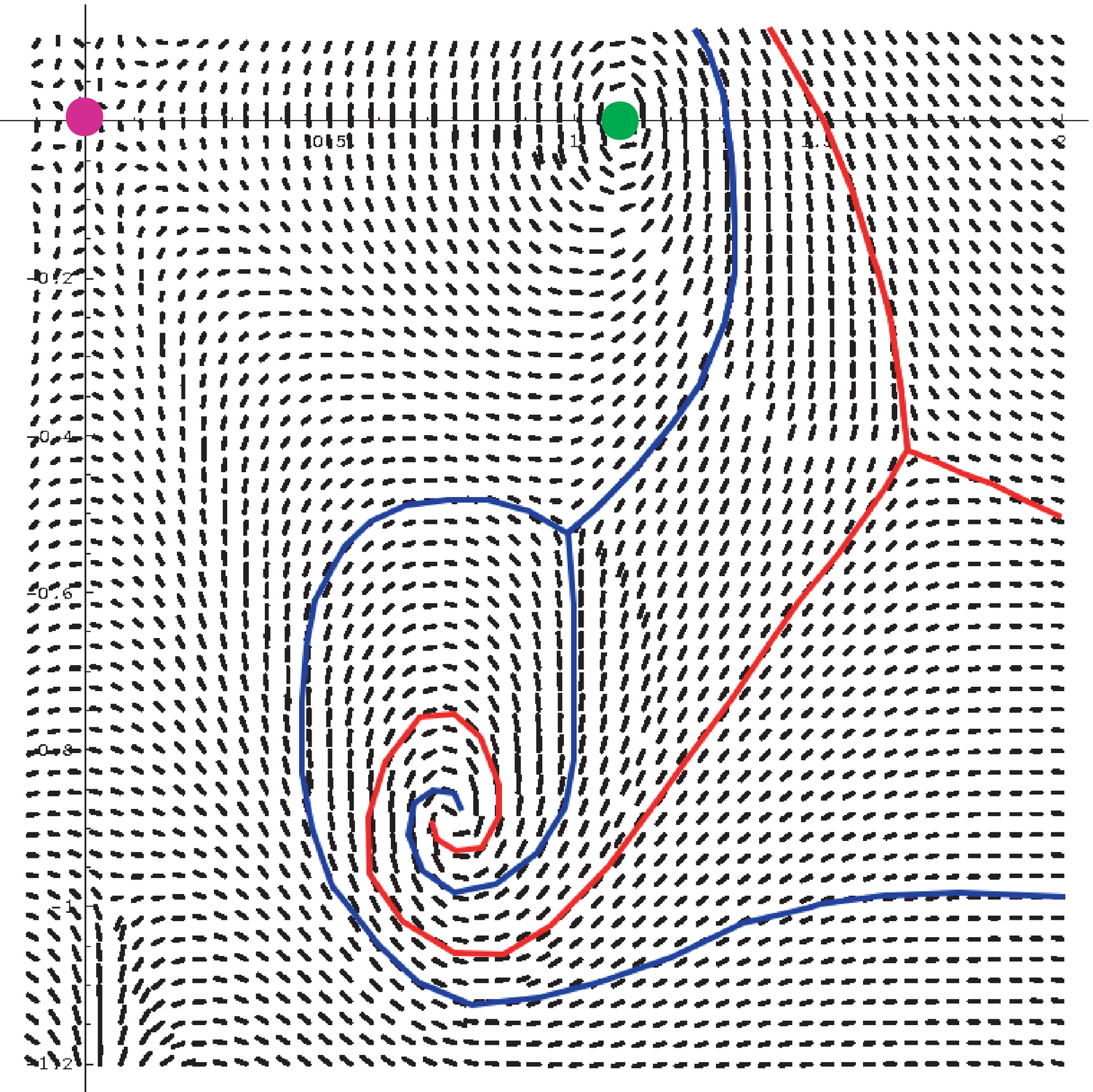} }
\ifig\plotsa{Same as above but for $u=2 e^{-i{\pi\ov 12}}$ and $u=2 e^{-i{\pi\ov 8}}$}{ \epsfxsize=3.5cm
\epsfbox{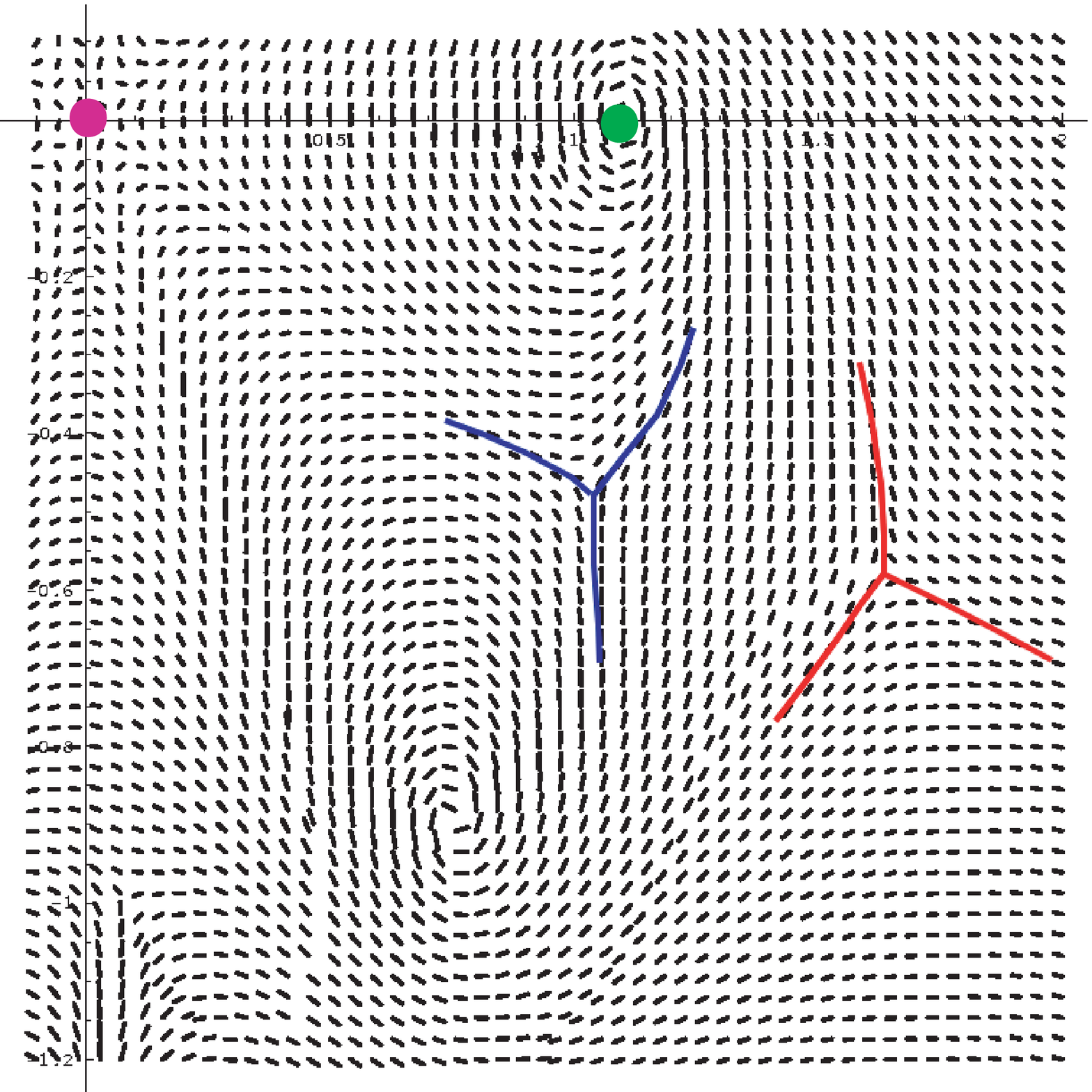} \epsfxsize=3.5cm \epsfbox{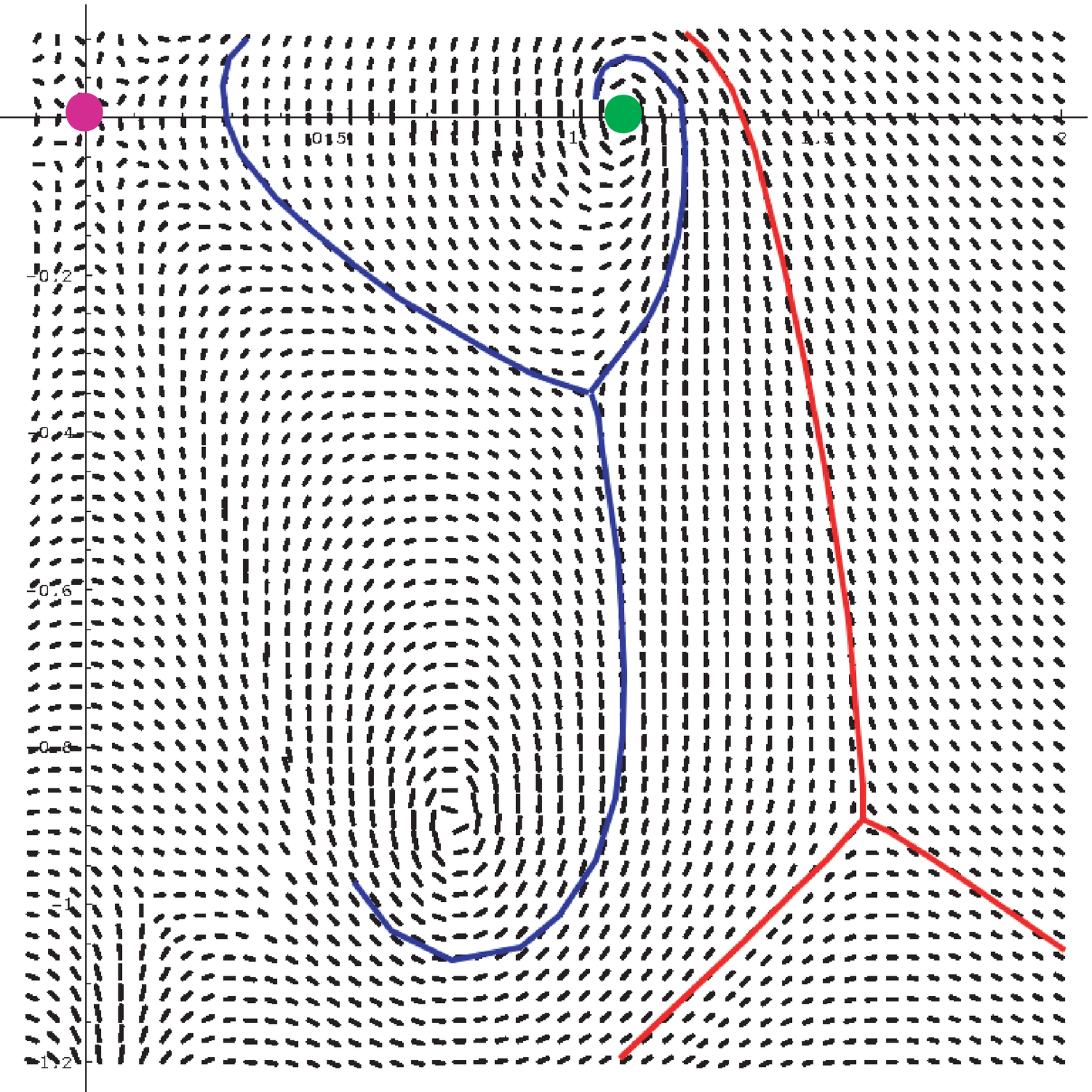}}
For $|u|$ large and some value of
$\arg(u)$ we have $Re(\ZZ(z_T,z_k))=0$ (that is figure $4$ above)
and the Stokes lines configuration in the $z$ plane is easiest to picture:

{\epsfxsize=7cm\epsfbox{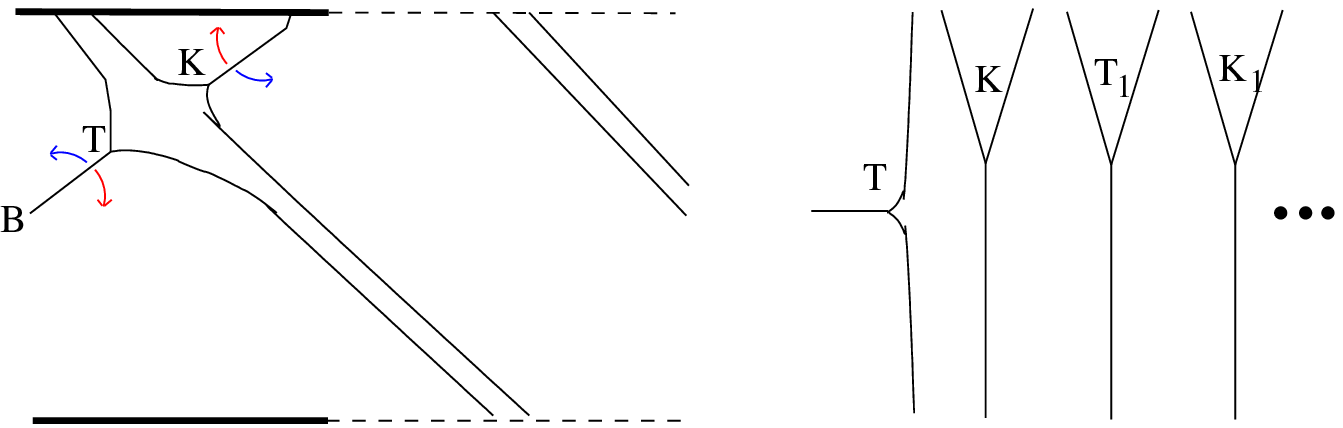}}

Also pictured is a schematic representation of the active region.
In this case too an infinite succession $T_n$ and $K_n$ of images
of the turning points is active.

Given the configuration of turning points we have to apply the
connection formula at $T\;K\;T_1\;K_1\;T_2\;K_2\;..$
counterclockwise. Proceeding as before we will have for $z$ in the
active region between $T$ and $K$

$$
g_{\om p}(z)\sim  \kappa(z,u)^{-\ha} A (e^{\nu
\ZZ({z_T},z)}+i e^{-\nu \ZZ({z_T},z)})
$$
where $A$ is defined as in \firss .
Then by using the connection formula at $K$ we get:
$$g_{\om p}(z)\sim  \kappa(z,u)^{-\ha}A\big(e^{\nu \ZZ({z_T},z)}+i e^{-\nu
\ZZ({z_T},z)}(1+e^{2\nu \ZZ({z_T},z_K)})\big)
$$
By repeating the steps for $T_1$ and $K_1$ we obtain:
$$
g_{\om p}(z) \sim \kappa(z,u)^{-\ha}A\big(e^{\nu
\ZZ({z_T},z)}+i e^{-\nu \ZZ({z_T},z)}(1+e^{2\nu
\ZZ({z_T},z_K)})(1+e^{2\nu \ZZ({z_T},z_{T_1})})\big)
$$
so that for $z\rightarrow +\infty$ taking into account the periodic succession of turning points we obtain
$$
g_{\om p}(z) \sim {A\ov i \nu u}\le(e^{i \nu u z+\gamma}+i e^{-i \nu u z-\gamma}(1+e^{2\nu \ZZ({z_T},z_K)}){2
e^{\nu u \beta/2}\ov \sinh( \nu u \beta/2)}\ri).
$$
with $\gamma$ defined by \defgamma .

From this formula it is easy to see that as the two turning points exchange dominance the exponential factor in $(1+e^{2\nu \ZZ({z_T},z_K)})$ is just a phase and the location of the quasinormal modes can be found by solving $$\ZZ({z_T},z_K)= i{\pi\ov 2\nu}(1+2n),\;\;\;n=0,1,2,\cdots$$

When $u$ is tilted toward the real axis $T$ and $K$ move in the
direction of the red arrows and $\exp(\ZZ({z_T},z_K))\ll 1$
For $u$ tilted towards the imaginary axis $T$ and $K$ move in the
direction of the blue arrow and $\exp(\ZZ({z_T},z_K))\gg1$.

\subsec{ $k\neq 0 $}

For $k\in R $ and small there are new solutions to $f(r)(1+{k^2\ov r})=0$ lying near the
singularity but the turning points giving the dominant contribution to $g_{\om p}(z)$ are the same as for
$k=0$ except for the fact that their position will depend continuously on $k$. The WKB approximations
to $g_{\om p}(z)$ are the same as before except that the integral defining $\ZZ$ has to be
evaluated at finite $k$.

\ifig\potg{The left figure is a schematic plot for the potential
$V(z)$ for $k > k_c$. On the right is a simplified scheme of the active region for $Im(u)<0\;\;\;Re(u)>0$} {\epsfxsize=7cm \epsfbox{pot.eps} \epsfxsize=7cm \epsfbox{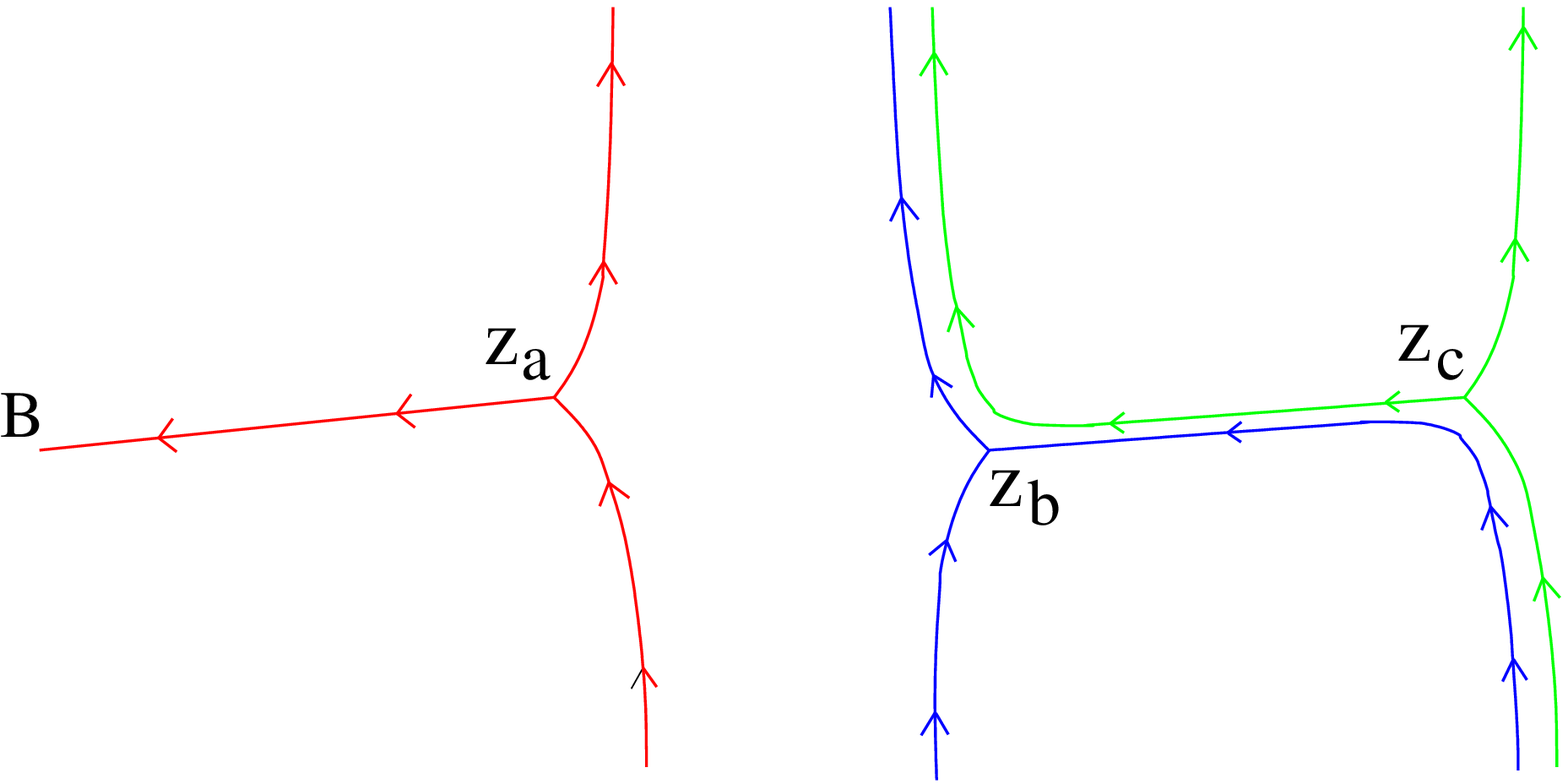}}

However as described in section 5.2 for finite $\mu$ when $k$
becomes greater than a critical value $k_c(\mu)$ the potential on
the real $z$ axis changes from being monotonically decreasing to
the one depicted in \potg\ . There is a region between $z_{min}$
and $z_{max}$ for which the potential is an increasing function of
$z$.

The resulting pattern of Stokes lines for $u$ close to the real axis and between $u_{min}$ and $u_{max}$ can be quite complex. However if we ignore the complications arising from the periodic structure of
the turning point in the $z$ plane we can picture the active region for $0<u_{min}<Re(u)<u_{max}$ and $u$ with a small negative imaginary part as in \potg.

This configuration of active turning points gives for the coefficient of $e^{-i \nu u z}$
in the WKB expression for $g_{\om p}(z)$ as $z\rightarrow \infty$:
\eqn\contfink{1+e^{2\nu \ZZ(z_a,z_c)}+e^{2\nu \ZZ(z_b,z_c)}=1+e^{2\nu \ZZ(z_b,z_c)}\le(1+e^{2\nu \ZZ(z_a,z_b)}\ri)}
Notice that with the choice of signs implied by the arrows in \potg\   $|e^{2\nu \ZZ(z_a,z_b)}|=1+\delta$ with $\delta \ll 1$ while $|e^{2\nu \ZZ(z_b,z_c)}|=\Lambda\gg 1$. So whenever \eqn\turnim{2\nu \ZZ(z_a,z_b)\approx i \pi(1+2n),\;\;\;n=0,1,2,\cdots} \contfink\ is zero as long as $\Lambda \delta=1$
An explicit expression for $\delta$ can be given in terms of the oscillation period $T(z_a,z_b)$ between $z_a$ and $z_b$ of a classical particle:
\eqn\defdelta{\delta= -2 \nu  u Im(u) \left|\int_{z_a}^{z_b} \kappa(u,k;z)^{-1}dz\right|= -\nu u Im(u) T(z_a,z_b)}
Therefore we get for the distance of the quasinormal modes from the real $u$ axis:
\eqn\deltau{Im(u)=-\le(\nu u\;T(z_a,z_b)\ri)^{-1} e^{-2\nu \ZZ(z_b,z_c)}}
As $u\rightarrow u_{max}$ the imaginary part of the quasinormal frequencies increases as $\Lambda$ decreases. The mode lines diverge from the real axis and for $u>u_{max}$ they join with those determined in the previous subsection as for large $u$ the non-monotonicity of the potential becomes less and less important.

\appendix{B}{Tortoise coordinate}

In this appendix the analytic structure of the function $r(z)$  will be described in some detail
and we will present its determination used in appendix A.

First recall the definition of the tortoise coordinate. \eqn\tortc {z(r)=\int_{r}^{\infty}dr'{1\ov
f(r')}} where $f(r)=r^2+1-{\mu\ov r^{d-2}}$.

The solutions of $f(r)=0$ will be denoted in the following by $r_i\;\;\;i=0,1,..,d-1$. In particular
$r_0$ is the positive real solution corresponding to the horizon. Also the residues at $r_i$ in the
integral for $z(r)$ will be denoted as $Res(r_i)={\beta_i\ov4 \pi}$. Then $\beta_0=\beta$ the
inverse temperature of the black hole.

In order to obtain a definite value of $z(r)$ we have to specify a particular contour connecting
$\infty$ to $r$. Suppose that one particular contour $\CC$ is chosen and the value $z_{\CC}(r)$ is
obtained. All the possible values of $z$ corresponding to the same point will be $z(r)=z_{\CC}(r)+2
i\sum_{i=0}^{d-1}\alpha_i \beta_i$ where $\alpha_i\in \IZ$.

We can give a unique prescription for $z(r)$ by imposing that the contour $\CC$ cannot cross
a set of lines starting from the $r_i$. For example we can take these lines as propagating from $0$
to the $r_i$ radially but this is only one of many possible choices \foot{note that this choice
works due to the fact that the sum of the residues is 0.}. These lines will then be branch cuts
for the function $z(r)$.

Some properties of the function $z(r)$ which will be useful in the following are:
\item{1} The behavior for $r\rightarrow \infty$ is $z(r)\sim {1\ov
r}$
\item{2} The behavior for $r\sim r_i$ is $z(r)\sim
-{\beta_i\ov 4 \pi}\log(r-r_i)$.
\item{3} The behavior for $r\sim 0$ is $z\sim {r^{d-1}\ov
(d-1)\mu}+z(0)$ where $z(0)$ is any one of its possible values. Viceversa $r\sim ((d-1)\mu
(z-z(0)))^{1\ov d-1}$.

From the last item in the list above we see that the function $r(z)$ is not single valued. The
following is the description of the particular determination of this function that is used in appendix A.
Notice that for $r\sim r_0$ the function $r(z)=r_0+\exp({4\pi z\ov \beta})$ is
periodic with period $i{\beta\ov 2}$. In fact the lines of constant real part for $z(r)$ around
$r_0$ are topologically $S_1$ around $r_0$. This is valid only for $Re(z)$ greater than ${\tilb \ov
4}=\int_0^{\infty} dr P[f(r)^{-1}]$ where $P$ denotes the principal part at the horizon. At this value of $Re(z)$ the line passes at $r=0$ for $z={\tilb \ov 4}+ i{\beta\ov
2}(n+\ha)\;\;\;n\in \IZ$. These points are then branch points for the function $r(z)$. We will
choose the branch cuts propagating from them to have constant imaginary part and to go towards
$Re(z)=-\infty$. By following this prescription for all branch points we find a determination of
$r(z)$ such that $r(z+i{\beta\ov 2})=r(z)$. With this choice of determination for $Re(z)>0$ the
only branch points are those described: $z={\tilb \ov 4}+ i{\beta\ov 2}(n+\ha)\;\;\;n\in \IZ$ .

\listrefs

\end
\end